\documentclass[conference,compsoc]{IEEEtran}

\usepackage[T1]{fontenc}
\usepackage[utf8]{inputenc}
\usepackage{graphicx}
\usepackage{booktabs}
\usepackage{array}
\usepackage{multirow}
\usepackage{amsmath,amssymb}
\usepackage{xcolor}
\usepackage{colortbl}
\usepackage{tikz}
\usepackage{nicematrix} 
\usetikzlibrary{positioning,arrows.meta,fit,backgrounds,calc}
\usepackage{pifont}
\usepackage[scaled=0.92]{inconsolata} 
\usepackage{enumitem}
\usepackage{balance}
\usepackage[hidelinks,breaklinks]{hyperref}
\usepackage{cleveref}

\usepackage[skins,breakable]{tcolorbox}

\definecolor{palPlum}{HTML}{34183E}
\definecolor{palBlue}{HTML}{4D779B}
\definecolor{palGray}{HTML}{585D5E}
\definecolor{palMagenta}{HTML}{82093B}
\definecolor{palRose}{HTML}{C45C69}
\definecolor{palRed}{HTML}{CD3B42}
\definecolor{palAmber}{HTML}{FFC04D}
\definecolor{palTeal}{HTML}{2E8B7F}  

\newcounter{takeaway}
\newcounter{openproblem}
\newcounter{finding}

\newtcolorbox{takeaway}[1][]{
  breakable, enhanced, sharp corners,
  before skip=5pt, after skip=5pt,
  colback=palBlue!7, colframe=palBlue, boxrule=0.5pt,
  colbacktitle=palBlue, coltitle=white,
  left=4pt,right=4pt,top=3pt,bottom=3pt,
  fonttitle=\bfseries\footnotesize,
  title={Takeaway~\refstepcounter{takeaway}\thetakeaway#1}}

\newtcolorbox{openproblem}[1][]{
  breakable, enhanced, sharp corners,
  before skip=5pt, after skip=5pt,
  colback=palPlum!6, colframe=palPlum, boxrule=0.5pt,
  colbacktitle=palPlum, coltitle=white,
  left=4pt,right=4pt,top=3pt,bottom=3pt,
  fonttitle=\bfseries\footnotesize,
  title={Open Problem~\refstepcounter{openproblem}\theopenproblem#1}}

\newtcolorbox{finding}[1][]{
  breakable, enhanced, sharp corners,
  before skip=5pt, after skip=5pt,
  colback=palMagenta!6, colframe=palMagenta, boxrule=0.5pt,
  colbacktitle=palMagenta, coltitle=white,
  left=4pt,right=4pt,top=3pt,bottom=3pt,
  fonttitle=\bfseries\footnotesize,
  title={Finding~\refstepcounter{finding}\thefinding#1}}

\newsavebox{\hbfullbox}\newsavebox{\hbhalfbox}\newsavebox{\hbnonebox}
\sbox{\hbfullbox}{\tikz[baseline=-0.55ex]\fill (0,0) circle (0.55ex);}
\sbox{\hbhalfbox}{\begin{tikzpicture}[baseline=-0.55ex]%
  \fill (0,0) -- (90:0.55ex) arc (90:270:0.55ex) -- cycle;%
  \draw[line width=0.4pt] (0,0) circle (0.55ex);\end{tikzpicture}}
\sbox{\hbnonebox}{\tikz[baseline=-0.55ex]\draw[line width=0.4pt] (0,0) circle (0.55ex);}
\newcommand{\cmark}{\usebox{\hbfullbox}}   
\newcommand{\pmark}{\usebox{\hbhalfbox}}   
\newcommand{\xmark}{\usebox{\hbnonebox}}   

\newcommand{\stage}[1]{\textsc{#1}}

\newcommand{\myparagraph}[1]{\vspace{2pt}\noindent\textbf{#1}}

\newif\ifshowedits
\showeditsfalse
\ifshowedits
  \newcommand{\edited}[1]{\textcolor{orange!85!black}{#1}}
\else
  \newcommand{\edited}[1]{#1}
\fi

\usepackage{listings}
\definecolor{diffaddfg}{HTML}{1E6B2A}   
\definecolor{diffremfg}{HTML}{B02A33}   
\definecolor{diffcmt}{HTML}{6A6F70}     
\definecolor{codeframe}{HTML}{4D779B}   
\lstdefinestyle{kdiff}{
  basicstyle=\ttfamily\footnotesize,
  columns=fullflexible, keepspaces=true, showstringspaces=false,
  breaklines=true, breakatwhitespace=true, breakindent=10pt,
  frame=l, framerule=1.6pt, rulecolor=\color{codeframe},
  framesep=6pt, xleftmargin=9pt, xrightmargin=3pt,
  aboveskip=5pt, belowskip=3pt,
  morecomment=[f][\color{diffaddfg}]{+},
  morecomment=[f][\color{diffremfg}]{-},
  morecomment=[f][\color{diffcmt}]{@@},
}

\begin{document}
\bstctlcite{BSTcontrol}

\title{SoK: From Crash to Patch: Systematizing the Operating Systems Kernel \\Bug Lifecycle}

\author{
  \IEEEauthorblockN{
    Luyao Bai\IEEEauthorrefmark{1},
    Gengda She\IEEEauthorrefmark{2},
    Kenan Alghythee\IEEEauthorrefmark{1}, 
    Hang Zhang\IEEEauthorrefmark{2}, and
    Xiaoguang Wang\IEEEauthorrefmark{1}
  }
  \IEEEauthorblockA{
    \IEEEauthorrefmark{1}University of Illinois Chicago\qquad
    \IEEEauthorrefmark{2}Indiana University Bloomington
  }
}

\maketitle

\begin{abstract}
Automated kernel bug discovery has advanced rapidly. Continuous fuzzing and static analysis systems, such as syzbot, now expose Linux kernel bugs at a scale that downstream processes struggle to absorb. Yet a crash report is only the beginning. Before a bug is eliminated, it must be triaged, understood, patched, validated, reviewed, integrated, and often backported. These later stages remain far less automated, creating a persistent gap between bug discovery and patch deployment.

This SoK systematizes the Linux kernel bug lifecycle from discovery to deployment. We organize prior work and production systems into five stages: discovery, triage, patch generation, patch validation, and integration. We explain the resulting automation gradient through kernel-specific challenges such as concurrency, implicit invariants, cross-syscall state, hardware dependence, lack of fault isolation, and architecture/configuration multiplicity. We further ground the analysis in a measurement of real syzbot-fixed bugs. The data shows that the crash-to-patch gap is not merely a backlog of unfixed reports but a structural failure mode of the repair pipeline: even after being fixed, bugs often remain open for weeks, require review-driven patch revisions, or lack reproducers that current repair and validation systems assume. This exposes a mismatch between where kernel-security automation is mature and where bug closure actually breaks down. These findings expose a deeper mismatch: today's repair and validation techniques often assume reliable reproducers, localized root causes, and checkable correctness oracles, yet these are precisely the artifacts missing from many real kernel bug reports. Closing the crash-to-patch gap, therefore, requires treating such artifacts as outputs to be produced, not prerequisites to be assumed.
\end{abstract}


\section{Introduction}
\label{sec:intro}

The operating system (OS) kernel is the largest and most privileged trusted
computing base on virtually every device. The Linux kernel alone exceeds 30
million lines of code, integrates thousands of patches per release, and is
maintained by a loosely coordinated community of volunteers and corporate
contributors. A single memory-safety or concurrency defect can compromise the
entire system, and memory-safety errors continue to account for roughly 70\% of
serious vulnerabilities in large C/C++ codebases including the
kernel~\cite{memsafety}. Securing the kernel is therefore not a single problem
but a pipeline of problems: a bug must be found, understood and triaged,
repaired and validated, and finally reviewed, integrated into a constantly
moving codebase, and backported to the stable trees.

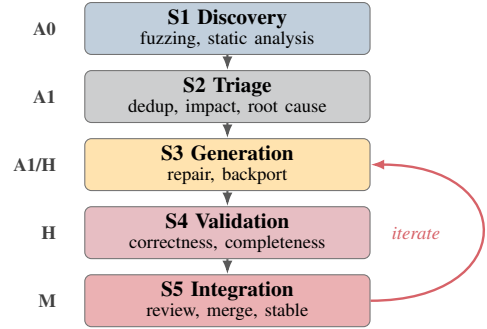
\begin{figure}[t]
  \centering
\begin{tikzpicture}[
    font=\footnotesize,
    stage/.style={rectangle, rounded corners=2.5pt, draw=black!55, line width=0.4pt,
      text width=3.6cm, align=center, minimum height=6mm, inner sep=2.5pt},
    badge/.style={font=\scriptsize\bfseries, text=black!75, anchor=east},
    arr/.style={-{Latex[length=1.8mm]}, line width=0.8pt, black!65},
  ]

  \node[stage, fill=palBlue!35]  (s1) {\textbf{S1 Discovery}\\[-1.5pt]\scriptsize fuzzing, static analysis};
  \node[stage, fill=palGray!30, below=2.0mm of s1] (s2) {\textbf{S2 Triage}\\[-1.5pt]\scriptsize dedup, impact, root cause};
  \node[stage, fill=palAmber!45, below=2.0mm of s2] (s3) {\textbf{S3 Generation}\\[-1.5pt]\scriptsize repair, backport};
  \node[stage, fill=palRose!40, below=2.0mm of s3] (s4) {\textbf{S4 Validation}\\[-1.5pt]\scriptsize correctness, completeness};
  \node[stage, fill=palRed!40, below=2.0mm of s4] (s5) {\textbf{S5 Integration}\\[-1.5pt]\scriptsize review, merge, stable};

  \foreach \a/\b in {s1/s2,s2/s3,s3/s4,s4/s5}{\draw[arr] (\a) -- (\b);}

  \node[badge] at ([xshift=-2.5mm]s1.west) {A0};
  \node[badge] at ([xshift=-2.5mm]s2.west) {A1};
  \node[badge] at ([xshift=-2.5mm]s3.west) {A1/H};
  \node[badge] at ([xshift=-2.5mm]s4.west) {H};
  \node[badge] at ([xshift=-2.5mm]s5.west) {M};

  \draw[arr, palRed!80, line width=0.9pt]
    (s5.east) to[out=0, in=0, looseness=2.8] (s3.east);
  \node[font=\scriptsize\itshape, text=palRed!80]
    at ([xshift=6mm]s4.east) {iterate};

\end{tikzpicture}
  \caption{The OS kernel bug lifecycle in five stages, from
  \stage{Discovery} to \stage{Integration}, with the automation level of each
  stage (A0 to M).}
  \label{fig:pipeline}
\end{figure}

The first stage of this pipeline has been transformed. Coverage-guided kernel
fuzzing now runs continuously at scale: Google's syzbot fuzzes mainline and
linux-next around the clock and has reported tens of thousands of bugs, and
static analysis has scaled to the whole kernel. Continuous fuzzing and static
analysis now report candidate bugs faster than the downstream pipeline can
process and close them; the bottleneck is no longer finding more bugs but
clearing the backlog of bugs already found.

The remaining stages have not scaled with it. Everything after discovery is still predominantly manual work for an overworked maintainer population~\cite{zhou2017,tan2020}: a maintainer must triage a crash into an actionable root cause, judge its security impact, write a patch preserving kernel invariants, show that it neither regresses nor partially fixes the bug, and shepherd it through mailing-list review and stable backporting. The visible symptom is a growing backlog, with fix latency varying enormously across subsystems. We refer to the distance between an automatically discovered crash and a deployed, validated fix as the crash-to-patch gap.

Most recently, large language models (LLMs) have been applied across the
pipeline in four distinct roles: as artifact generators (syscall
specifications, static checkers, candidate patches, review comments), as
classifiers/judges (severity, patch correctness), as agents that drive a repair
or integration workflow, and, aspirationally, as reasoning engines for root
cause and fix completeness. Yet this adoption is uneven and largely
unsystematized, with strong evidence for the first two roles and thin evidence
for the last.

Existing systematizations address slices of this pipeline: kernel-fuzzing
surveys~\cite{fuzzsurvey} cover discovery, the SoK on automated vulnerability
repair~\cite{sokavr} covers user-space repair, and the SoK on kernel
hardening~\cite{sokhardening} covers exploitation and mitigation, orthogonal to
bug management. None unifies discovery with triage, repair, validation, and the
socio-technical integration process, and none treats the kernel's defining
characteristic, a pipeline automated at the front and manual at the back. We
argue this end-to-end, kernel-specific view is exactly what is needed to direct
the field's next decade of effort.

Overall, this paper makes the following contributions:
\begin{itemize}[leftmargin=1.2em,itemsep=2pt]
  \item We define the \textit{OS kernel bug lifecycle} as a five-stage pipeline
  and use it to systematize 140 research efforts and production systems against
  a common set of dimensions (\Cref{sec:overview}--\ref{sec:integration}).

  \item We articulate the \textit{automation gradient} as the field's defining
  structural property and tie it to \textit{six kernel-specific challenges}
  (\Cref{sec:challenges}) whose difficulty is back-loaded onto the later
  stages, explaining where LLMs have and have not closed the gap.

  \item We conduct a \textit{longitudinal measurement of the crash-to-patch
  gap} over 6{,}946 fixed syzbot bugs, decomposing fix latency into pipeline
  segments\edited{, locating where the delay sits,} and \edited{computing} a repair-readiness
  score (\Cref{sec:measurement}).

  \item We build a \textit{coverage-gap matrix} mapping existing techniques onto
  (bug class $\times$ lifecycle stage), exposing combinations no current tool
  addresses, and distill \textit{takeaways} and \textit{open problems} per
  stage.

  \item We release our dataset, classification, and analysis scripts as a public
  artifact for reproducibility and continued community curation.
\end{itemize}

We focus on the kernel and the management of bugs from
discovery to deployment, treating user-space techniques (general APR,
code-review research) as contrast to highlight what is genuinely
kernel-specific. Exploitation and runtime hardening are out of scope, as they
concern defending against bugs rather than fixing them and are covered by a
complementary SoK~\cite{sokhardening}.

\section{Methodology and Scope}
\label{sec:methodology}
\begin{figure*}[!t]
  \centering
\tikzset{
  pillS1/.style={rectangle, rounded corners=2.5pt, draw=palBlue!75, fill=palBlue!13, line width=0.35pt, inner xsep=2.2pt, inner ysep=1.1pt, font=\scriptsize, text=black},
  pillS2/.style={rectangle, rounded corners=2.5pt, draw=palTeal!75, fill=palTeal!13, line width=0.35pt, inner xsep=2.2pt, inner ysep=1.1pt, font=\scriptsize, text=black},
  pillS3/.style={rectangle, rounded corners=2.5pt, draw=palAmber!75, fill=palAmber!13, line width=0.35pt, inner xsep=2.2pt, inner ysep=1.1pt, font=\scriptsize, text=black},
  pillS4/.style={rectangle, rounded corners=2.5pt, draw=palRose!75, fill=palRose!13, line width=0.35pt, inner xsep=2.2pt, inner ysep=1.1pt, font=\scriptsize, text=black},
  pillS5/.style={rectangle, rounded corners=2.5pt, draw=palRed!75, fill=palRed!13, line width=0.35pt, inner xsep=2.2pt, inner ysep=1.1pt, font=\scriptsize, text=black},
}
\renewcommand{\arraystretch}{1.10}
\setlength{\tabcolsep}{3pt}
\begin{NiceTabular}{@{}>{\centering\arraybackslash}m{0.085\textwidth}|>{\raggedright\arraybackslash}m{0.885\textwidth}@{}}[hlines, code-before={\cellcolor{palPlum}{1-1}\cellcolor{palBlue}{2-1}\cellcolor{palGray}{3-1}\cellcolor{palMagenta}{4-1}\cellcolor{palRose}{5-1}\cellcolor{palRed}{6-1}\cellcolor{palGray!70}{7-1}}]
{\color{white}\footnotesize\textbf{C1}\par\scriptsize Concurrency} & \tikz[baseline=(p.base)]\node[pillS1] (p) {Razzer\,\cite{razzer}}; \tikz[baseline=(p.base)]\node[pillS1] (p) {SegFuzz\,\cite{segfuzz}}; \tikz[baseline=(p.base)]\node[pillS1] (p) {Snowboard\,\cite{snowboard}}; \tikz[baseline=(p.base)]\node[pillS1] (p) {DCUAF\,\cite{dcuaf}}; \tikz[baseline=(p.base)]\node[pillS1] (p) {Double-Fetch\,\cite{doublefetch}}; \tikz[baseline=(p.base)]\node[pillS1] (p) {KRACE\,\cite{krace}}; \tikz[baseline=(p.base)]\node[pillS1] (p) {Snowcat\,\cite{snowcat}}; \tikz[baseline=(p.base)]\node[pillS1] (p) {UACatcher\,\cite{ma2023top}}; \tikz[baseline=(p.base)]\node[pillS1] (p) {DEADLINE\,\cite{xu2018precise}}; \\
{\color{white}\footnotesize\textbf{C2}\par\scriptsize Implicit invariants} & \tikz[baseline=(p.base)]\node[pillS1] (p) {Hydra\,\cite{hydra_sosp}}; \tikz[baseline=(p.base)]\node[pillS1] (p) {CountDown\,\cite{countdown}}; \tikz[baseline=(p.base)]\node[pillS1] (p) {DR.CHECKER\,\cite{drchecker}}; \tikz[baseline=(p.base)]\node[pillS1] (p) {K-Miner\,\cite{kminer}}; \tikz[baseline=(p.base)]\node[pillS1] (p) {CRIX\,\cite{crix}}; \tikz[baseline=(p.base)]\node[pillS1] (p) {LRSan\,\cite{lrsan}}; \tikz[baseline=(p.base)]\node[pillS1] (p) {UBITect\,\cite{ubitect}}; \tikz[baseline=(p.base)]\node[pillS1] (p) {K-MELD\,\cite{kmeld}}; \tikz[baseline=(p.base)]\node[pillS1] (p) {Goshawk\,\cite{goshawk}}; \tikz[baseline=(p.base)]\node[pillS1] (p) {STACK\,\cite{optsafe}}; \tikz[baseline=(p.base)]\node[pillS1] (p) {KNighter\,\ding{72}\,\cite{knighter}}; \tikz[baseline=(p.base)]\node[pillS1] (p) {Coccinelle\,\cite{coccinelle}}; \tikz[baseline=(p.base)]\node[pillS1] (p) {PATA\,\cite{li2022path}}; \tikz[baseline=(p.base)]\node[pillS1] (p) {KUBO\,\cite{liu2021kubo}}; \tikz[baseline=(p.base)]\node[pillS1] (p) {LLift\,\ding{72}\,\cite{li2024enhancing}}; \tikz[baseline=(p.base)]\node[pillS1] (p) {IMMI\,\cite{liu2024detecting}}; \tikz[baseline=(p.base)]\node[pillS1] (p) {MANTA\,\cite{yang2022making}}; \tikz[baseline=(p.base)]\node[pillS1] (p) {CRED\,\cite{cred}}; \tikz[baseline=(p.base)]\node[pillS1] (p) {CID\,\cite{cid}}; \tikz[baseline=(p.base)]\node[pillS1] (p) {PeX\,\cite{pex}}; \tikz[baseline=(p.base)]\node[pillS1] (p) {CheQ\,\cite{lu2019automatically}}; \tikz[baseline=(p.base)]\node[pillS1] (p) {Err-Spec\,\cite{dossche2024inference}}; \tikz[baseline=(p.base)]\node[pillS1] (p) {DEPA\,\cite{zhong2020inferring}}; \tikz[baseline=(p.base)]\node[pillS1] (p) {Uninit-Bin\,\cite{garmany2019static}}; \tikz[baseline=(p.base)]\node[pillS3] (p) {kGym\,\cite{kgym}}; \tikz[baseline=(p.base)]\node[pillS3] (p) {CrashFixer\,\ding{72}\,\cite{crashfixer}}; \tikz[baseline=(p.base)]\node[pillS3] (p) {PatchIsland\,\ding{72}\,\cite{patchisland}}; \tikz[baseline=(p.base)]\node[pillS3] (p) {Beyond-C2P\,\cite{beyondc2p}}; \tikz[baseline=(p.base)]\node[pillS3] (p) {ChatRepair\,\ding{72}\,{\tiny(U)}\,\cite{chatgpt162}}; \tikz[baseline=(p.base)]\node[pillS3] (p) {Learn2Fix\,{\tiny(U)}\,\cite{hitlapr}}; \tikz[baseline=(p.base)]\node[pillS3] (p) {RGym\,\ding{72}\,\cite{rgym2025}}; \tikz[baseline=(p.base)]\node[pillS3] (p) {RepairAgent\,\ding{72}\,{\tiny(U)}\,\cite{repairagent}}; \tikz[baseline=(p.base)]\node[pillS3] (p) {AutoCodeRover\,\ding{72}\,{\tiny(U)}\,\cite{autocoderover}}; \tikz[baseline=(p.base)]\node[pillS3] (p) {ThinkRepair\,\ding{72}\,{\tiny(U)}\,\cite{thinkrepair}}; \tikz[baseline=(p.base)]\node[pillS4] (p) {KLAUS\,\cite{klaus}}; \tikz[baseline=(p.base)]\node[pillS4] (p) {Patch-Impact\,{\tiny(U)}\,\cite{patchimpact}}; \tikz[baseline=(p.base)]\node[pillS4] (p) {LLM-judge\,\ding{72}\,{\tiny(U)}\,\cite{llmjudge}}; \tikz[baseline=(p.base)]\node[pillS4] (p) {Incomplete-Fixes\,\cite{incompletefix}}; \tikz[baseline=(p.base)]\node[pillS4] (p) {PVBench\,{\tiny(U)}\,\cite{pvbench}}; \tikz[baseline=(p.base)]\node[pillS5] (p) {LiveBench\,\cite{livebench}}; \\
{\color{white}\footnotesize\textbf{C3}\par\scriptsize Cross-call state} & \tikz[baseline=(p.base)]\node[pillS1] (p) {kAFL\,\cite{kafl}}; \tikz[baseline=(p.base)]\node[pillS1] (p) {Horus\,\cite{horus}}; \tikz[baseline=(p.base)]\node[pillS1] (p) {BoKASAN\,\cite{bokasan}}; \tikz[baseline=(p.base)]\node[pillS1] (p) {MoonShine\,\cite{moonshine}}; \tikz[baseline=(p.base)]\node[pillS1] (p) {HEALER\,\cite{healer}}; \tikz[baseline=(p.base)]\node[pillS1] (p) {MOCK\,\cite{mock}}; \tikz[baseline=(p.base)]\node[pillS1] (p) {ACTOR\,\cite{actor}}; \tikz[baseline=(p.base)]\node[pillS1] (p) {SyzVegas\,\cite{syzvegas}}; \tikz[baseline=(p.base)]\node[pillS1] (p) {Snowplow\,\cite{snowplow}}; \tikz[baseline=(p.base)]\node[pillS1] (p) {IMF\,\cite{imf}}; \tikz[baseline=(p.base)]\node[pillS1] (p) {SyzGen\,\cite{syzgen}}; \tikz[baseline=(p.base)]\node[pillS1] (p) {KSG\,\cite{ksg}}; \tikz[baseline=(p.base)]\node[pillS1] (p) {SyzDescribe\,\cite{syzdescribe}}; \tikz[baseline=(p.base)]\node[pillS1] (p) {KernelGPT\,\ding{72}\,\cite{kernelgpt}}; \tikz[baseline=(p.base)]\node[pillS1] (p) {JANUS\,\cite{janus}}; \tikz[baseline=(p.base)]\node[pillS1] (p) {Hydra\,\cite{hydra_sosp}}; \tikz[baseline=(p.base)]\node[pillS1] (p) {StateFuzz\,\cite{statefuzz}}; \tikz[baseline=(p.base)]\node[pillS1] (p) {HFL\,\cite{hfl}}; \tikz[baseline=(p.base)]\node[pillS1] (p) {SyzDirect\,\cite{syzdirect}}; \tikz[baseline=(p.base)]\node[pillS1] (p) {syzkaller\,\cite{syzkaller}}; \tikz[baseline=(p.base)]\node[pillS1] (p) {Bin-Cov\,\cite{liu2024leveraging}}; \tikz[baseline=(p.base)]\node[pillS1] (p) {SyzGen++\,\cite{chen2024syzgen++}}; \tikz[baseline=(p.base)]\node[pillS1] (p) {FuzzNG\,\cite{bulekov2023no}}; \tikz[baseline=(p.base)]\node[pillS1] (p) {SUTURE\,\cite{zhang2021statically}}; \tikz[baseline=(p.base)]\node[pillS1] (p) {BugLens\,\ding{72}\,\cite{li2025towards}}; \tikz[baseline=(p.base)]\node[pillS1] (p) {UAFX\,\cite{zhang2025statically}}; \tikz[baseline=(p.base)]\node[pillS2] (p) {Dup-reports\,\cite{dupreports}}; \tikz[baseline=(p.base)]\node[pillS2] (p) {SyzRetrospector\,\cite{syzretrospector}}; \tikz[baseline=(p.base)]\node[pillS2] (p) {AURORA\,\cite{aurora}}; \tikz[baseline=(p.base)]\node[pillS2] (p) {ARCUS\,\cite{arcus}}; \tikz[baseline=(p.base)]\node[pillS2] (p) {Igor\,\cite{igor}}; \\
{\color{white}\footnotesize\textbf{C4}\par\scriptsize Hardware} & \tikz[baseline=(p.base)]\node[pillS1] (p) {Unicorefuzz\,\cite{unicorefuzz}}; \tikz[baseline=(p.base)]\node[pillS1] (p) {Agamotto\,\cite{agamotto}}; \tikz[baseline=(p.base)]\node[pillS1] (p) {SyzDescribe\,\cite{syzdescribe}}; \tikz[baseline=(p.base)]\node[pillS1] (p) {DIFUZE\,\cite{difuze}}; \tikz[baseline=(p.base)]\node[pillS1] (p) {DR.FUZZ\,\cite{drfuzz}}; \tikz[baseline=(p.base)]\node[pillS1] (p) {PrIntFuzz\,\cite{printfuzz}}; \tikz[baseline=(p.base)]\node[pillS1] (p) {KextFuzz\,\cite{kextfuzz}}; \tikz[baseline=(p.base)]\node[pillS1] (p) {NTFUZZ\,\cite{ntfuzz}}; \tikz[baseline=(p.base)]\node[pillS1] (p) {DR.CHECKER\,\cite{drchecker}}; \tikz[baseline=(p.base)]\node[pillS1] (p) {Nyx\,\cite{nyx}}; \tikz[baseline=(p.base)]\node[pillS1] (p) {USBFuzz\,\cite{usbfuzz}}; \tikz[baseline=(p.base)]\node[pillS1] (p) {ReUSB\,\cite{reusb}}; \tikz[baseline=(p.base)]\node[pillS5] (p) {Fast-fixes\,\cite{fastfixes}}; \\
{\color{white}\footnotesize\textbf{C5}\par\scriptsize No isolation} & \tikz[baseline=(p.base)]\node[pillS1] (p) {Digtool\,\cite{digtool}}; \tikz[baseline=(p.base)]\node[pillS1] (p) {SCAD\,\cite{man2025scad}}; \tikz[baseline=(p.base)]\node[pillS2] (p) {SyzScope\,\cite{syzscope}}; \tikz[baseline=(p.base)]\node[pillS2] (p) {DiffCVSS\,\cite{diffcvss}}; \tikz[baseline=(p.base)]\node[pillS2] (p) {LLM-triage\,\ding{72}\,\cite{llmtriage}}; \tikz[baseline=(p.base)]\node[pillS2] (p) {Vuln-Prediction\,\cite{vulnpredict}}; \tikz[baseline=(p.base)]\node[pillS2] (p) {GREBE\,\cite{grebe}}; \tikz[baseline=(p.base)]\node[pillS2] (p) {KOOBE\,\cite{koobe}}; \tikz[baseline=(p.base)]\node[pillS2] (p) {K-LEAK\,\cite{kleak}}; \tikz[baseline=(p.base)]\node[pillS2] (p) {FUZE\,\cite{fuze}}; \tikz[baseline=(p.base)]\node[pillS2] (p) {KEPLER\,\cite{kepler}}; \\
{\color{white}\footnotesize\textbf{C6}\par\scriptsize Arch/multi-tree} & \tikz[baseline=(p.base)]\node[pillS1] (p) {SyzRisk\,\cite{syzrisk}}; \tikz[baseline=(p.base)]\node[pillS1] (p) {Coccinelle\,\cite{coccinelle}}; \tikz[baseline=(p.base)]\node[pillS1] (p) {IncreLux\,\cite{zhai2022progressive}}; \tikz[baseline=(p.base)]\node[pillS1] (p) {Kconfig\,\cite{oh2021finding}}; \tikz[baseline=(p.base)]\node[pillS2] (p) {DiffCVSS\,\cite{diffcvss}}; \tikz[baseline=(p.base)]\node[pillS2] (p) {PatchScout\,\cite{patchscout}}; \tikz[baseline=(p.base)]\node[pillS2] (p) {DisPatch\,\cite{dispatch}}; \tikz[baseline=(p.base)]\node[pillS2] (p) {SPAIN\,\cite{spain}}; \tikz[baseline=(p.base)]\node[pillS2] (p) {SyzBridge\,\cite{syzbridge}}; \tikz[baseline=(p.base)]\node[pillS2] (p) {SPI\,\cite{spi}}; \tikz[baseline=(p.base)]\node[pillS3] (p) {Collateral-Evol.\,\cite{collateral}}; \tikz[baseline=(p.base)]\node[pillS3] (p) {FixMorph\,\cite{fixmorph}}; \tikz[baseline=(p.base)]\node[pillS3] (p) {PatchNet\,\cite{patchnet}}; \tikz[baseline=(p.base)]\node[pillS3] (p) {PatchScope\,\cite{patchscope}}; \tikz[baseline=(p.base)]\node[pillS3] (p) {Patch-porting\,\cite{patchporting}}; \tikz[baseline=(p.base)]\node[pillS4] (p) {PDiff\,\cite{pdiff}}; \tikz[baseline=(p.base)]\node[pillS4] (p) {PS3\,\cite{ps3}}; \tikz[baseline=(p.base)]\node[pillS5] (p) {Patch-Me-If-Can\,\cite{patchmeifyoucan}}; \tikz[baseline=(p.base)]\node[pillS5] (p) {CVE-coord.\,\cite{cnacoord}}; \tikz[baseline=(p.base)]\node[pillS5] (p) {Seamless-upd.\,\cite{seamless}}; \\
{\color{white}\footnotesize\textbf{Socio-}\par\footnotesize\textbf{technical}} & \tikz[baseline=(p.base)]\node[pillS5] (p) {Tufano'21\,\ding{72}\,{\tiny(U)}\,\cite{tufano2021}}; \tikz[baseline=(p.base)]\node[pillS5] (p) {CodeReviewer\,\ding{72}\,{\tiny(U)}\,\cite{codereviewer}}; \tikz[baseline=(p.base)]\node[pillS5] (p) {AUGER\,{\tiny(U)}\,\cite{auger}}; \tikz[baseline=(p.base)]\node[pillS5] (p) {ReviewBench\,\ding{72}\,{\tiny(U)}\,\cite{llmreviewbench}}; \tikz[baseline=(p.base)]\node[pillS5] (p) {DPO-f+\,\ding{72}\,{\tiny(U)}\,\cite{dpofplus}}; \tikz[baseline=(p.base)]\node[pillS5] (p) {ReviewStudy\,\ding{72}\,{\tiny(U)}\,\cite{llmreviewstudy}}; \tikz[baseline=(p.base)]\node[pillS5] (p) {Bacchelli\,{\tiny(U)}\,\cite{bacchelli2013}}; \tikz[baseline=(p.base)]\node[pillS5] (p) {McIntosh\,{\tiny(U)}\,\cite{mcintosh2016}}; \tikz[baseline=(p.base)]\node[pillS5] (p) {Ruangwan\,{\tiny(U)}\,\cite{ruangwan2019}}; \tikz[baseline=(p.base)]\node[pillS5] (p) {Goncalves\,{\tiny(U)}\,\cite{goncalves2022}}; \tikz[baseline=(p.base)]\node[pillS5] (p) {Incivility\,\cite{ferreira2021}}; \tikz[baseline=(p.base)]\node[pillS5] (p) {Patch-comm.\,\cite{tan2019communicate}}; \tikz[baseline=(p.base)]\node[pillS5] (p) {Zhou\,\cite{zhou2017}}; \tikz[baseline=(p.base)]\node[pillS5] (p) {Tan\,\cite{tan2020}}; \tikz[baseline=(p.base)]\node[pillS5] (p) {Onboarding\,{\tiny(U)}\,\cite{onboarding}}; \tikz[baseline=(p.base)]\node[pillS5] (p) {Rust-for-Linux\,\cite{rust4linux}}; \tikz[baseline=(p.base)]\node[pillS5] (p) {Women-in-OSS\,{\tiny(U)}\,\cite{womenoss}}; \tikz[baseline=(p.base)]\node[pillS5] (p) {Jiang\,\cite{jiang2013}}; \tikz[baseline=(p.base)]\node[pillS5] (p) {Disclosure\,{\tiny(U)}\,\cite{disclosure}}; \tikz[baseline=(p.base)]\node[pillS5] (p) {Coverity-alerts\,{\tiny(U)}\,\cite{coverityalerts}}; \tikz[baseline=(p.base)]\node[pillS5] (p) {Patchwork\,\cite{patchwork}}; \\
\end{NiceTabular}
\\[3pt]
{\scriptsize \textbf{Pill color = lifecycle stage:}~ \tikz\node[pillS1]{\textbf{S1}}; Discovery \quad \tikz\node[pillS2]{\textbf{S2}}; Triage \quad \tikz\node[pillS3]{\textbf{S3}}; Generation \quad \tikz\node[pillS4]{\textbf{S4}}; Validation \quad \tikz\node[pillS5]{\textbf{S5}}; Integration.
\quad \ding{72}~= LLM-based \quad (U)~= user-space contrast.
A paper that confronts several challenges appears in several rows.}

  \caption{Classification of the surveyed systems by kernel challenge
  (rows, C1--C6) and lifecycle stage (pill color, S1--S5).}
  \label{fig:classification}
\end{figure*}

\subsection{Paper Selection}

We assembled our corpus in three steps. \emph{Seed search}: we queried DBLP,
Google Scholar, and the proceedings of top security (S\&P, USENIX Security, CCS,
NDSS), systems (OSDI, SOSP, EuroSys, ATC, ASPLOS), and software-engineering
(ICSE, FSE, ASE, ISSTA) venues, combining kernel with bug discovery, vulnerability repair, patch
generation, patch correctness, backporting, and code review, over a primary
window of 2015--2026 (the rise of continuous kernel fuzzing through the LLM
era), admitting seminal earlier work such as the \emph{Faults in Linux}
studies~\cite{faultslinux10,faultslinux26} where it anchors a category.
\emph{Filtering}: we retained a paper if it (i) targets or substantially
evaluates on the OS kernel and (ii) contributes to at least one lifecycle stage,
admitting a bounded set of user-space and software-engineering papers as contrast
where they expose a kernel-specific gap, and excluding work targeting solely
exploitation or runtime hardening. \emph{Snowballing}: we chased citations
forward and backward until no new methodologically distinct work appeared
(two iterations). The final corpus totals 140 papers.

\subsection{The Lifecycle Lens}

We organize the corpus around the five lifecycle stages a kernel bug traverses
from existence to eradication (\Cref{fig:pipeline}): \stage{S1 Discovery}
exposes a latent defect as an observable failure (\Cref{sec:discovery});
\stage{S2 Triage \& Understanding} deduplicates, root-causes, and assesses
impact (\Cref{sec:triage}); \stage{S3 Patch Generation} synthesizes a candidate
fix, including backports (\Cref{sec:generation}); \stage{S4 Patch Validation}
establishes that the fix resolves the bug, preserves functionality, and is
complete (\Cref{sec:validation}); and \stage{S5 Integration} reviews, merges,
and ships it through the community process (\Cref{sec:integration}). Three kinds
of callout boxes thread the paper: Takeaways synthesize each stage, Open Problems
mark unresolved challenges, and Findings report our measurement results
(\Cref{sec:measurement}).

\subsection{The Automation Gradient}

Our unifying claim is structural. We assign each surveyed technique an automation
level on a four-point scale that distinguishes deployed from merely demonstrable
automation:
\begin{itemize}[leftmargin=1.4em,itemsep=1pt,topsep=2pt]
  \item \textbf{A0}: fully automatic in a production pipeline with no human in
  the loop (e.g., syzbot's continuous fuzzing).
  \item \textbf{A1}: fully automatic per input, but offline, per-bug, or a
  prototype outside any continuous pipeline (e.g., current LLM repair agents).
  \item \textbf{H}: human-guided, where the tool proposes and a human decides.
  \item \textbf{M}: manual best practice in which a tool merely assists (e.g.,
  mailing-list code review).
\end{itemize}
The distinction matters because much back-end ``automation'' is A1: it works in a
paper but has never been wired into the syzbot-scale flow, so it does not relieve
the production bottleneck. The modal level degrades monotonically across
S1$\rightarrow$S5, and only discovery reaches A0. \Cref{sec:overview} makes this
gradient precise, the per-stage sections substantiate it, and
\Cref{sec:measurement} shows its consequence in the wild as the crash-to-patch
gap.

\section{The Kernel Bug Lifecycle at a Glance}
\label{sec:overview}

\Cref{fig:pipeline} presents the five-stage lifecycle and the automation
gradient that is this paper's organizing thesis. We call the early, automated
stages (S1--S2) the front end and the later, \edited{human-in-the-loop} stages (S3--S5) the back
end: reading top to bottom, production-deployed automation (A0) thins out, only
discovery reaches it, and the later stages lean on offline prototypes (A1) and
human judgment (H, M).

Automation level is only one dimension per
stage. \Cref{tab:taxonomy} broadens this into the full framework we use
throughout, recording for each stage the artifacts it consumes and produces, its
dominant method, automation level, the oracle that defines when the stage is
``done'', the role LLMs play, and the challenges (\Cref{sec:challenges}) that
constrain it. Reading top to bottom, every column weakens together; the gradient
is this same decline, seen along five axes at once.

\begin{table*}[t]
  \centering
  \caption{A cross-cutting taxonomy of the kernel bug lifecycle,
  applied uniformly across all five stages.}
  \label{tab:taxonomy}
  \scriptsize
  \setlength{\tabcolsep}{5pt}
  \renewcommand{\arraystretch}{1.06}
  \begin{tabular}{@{}p{1.9cm} p{3.5cm} p{2.5cm} c p{3.0cm} p{2.6cm} c@{}}
    \toprule
    \textbf{Stage} & \textbf{Consumes $\rightarrow$ Produces} &
    \textbf{Dominant method} & \textbf{Auto.} &
    \textbf{Oracle (``done'')} & \textbf{Productive LLM role} &
    \textbf{Chal.} \\
    \midrule
    \stage{S1} Discovery &
    source/binary $\rightarrow$ crash + report &
    coverage-guided search, static analysis & A0 &
    \emph{strong}: crash / sanitizer trips &
    generator: specs, checkers & C1,C3,C4 \\

    \stage{S2} Triage &
    crash + report $\rightarrow$ root cause, severity, dedup &
    symbolic exec., static, learning & A1 &
    \emph{partial}: exploit primitive, dup match & judge: severity labels &
    C3,C5 \\

    \stage{S3} Generation &
    root cause $\rightarrow$ candidate patch &
    templates, transforms, LLM agents & A1/H &
    \emph{weak}: a single reproducer & agent: propose patch &
    C2,C6 \\

    \stage{S4} Validation &
    patch $\rightarrow$ correct/complete verdict &
    directed fuzzing, static, PoC & H &
    \emph{weak/none}: no spec for completeness & judge: correctness (unverified) &
    C2,C5 \\

    \stage{S5} Integration &
    patch $\rightarrow$ merged + backported fix &
    human review, social process & M &
    \emph{social}: maintainer acceptance & generator: review comments &
    C6 \\
    \bottomrule
  \end{tabular}
\end{table*}

\stage{Discovery} runs unattended and
produces bugs faster than they can be processed; \stage{Triage} automates
deduplication and impact but not root cause; \stage{Generation} proposes patches
at low accepted yields; \stage{Validation} is largely manual; and at
\stage{Integration} the limit is maintainer bandwidth itself. This gradient is
not an accident of effort allocation; \Cref{sec:challenges} argues it follows
from six kernel-specific properties whose difficulty falls most heavily on
exactly these back-end stages.

\section{Why Kernel Bugs Are Different}
\label{sec:challenges}
A natural objection to a kernel bug-lifecycle SoK is that it merely re-targets
user-space bug finding and automated program repair (APR). This section answers
that objection and supplies the analytical lens for the rest of the paper: six
cross-cutting properties of the OS kernel that shape every stage of its bug
lifecycle (\Cref{sec:ch-six}), each grounded in a representative merged fix
from our corpus (\Cref{sec:ch-wild}). These properties burden the pipeline
asymmetrically, falling far more heavily on the later stages than on discovery,
which aligns with and helps explain the automation gradient
(\Cref{sec:ch-asym}).

\subsection{Six Cross-Cutting Challenges}
\label{sec:ch-six}

\myparagraph{C1: Pervasive concurrency and weak memory ordering.} The kernel
executes concurrently on all CPUs, with preemption, interrupts, and RCU. Many
defects (data races, deadlocks, use-after-free via concurrent free) are
properties of a particular interleaving, not of an input, whereas most user-space
APR and fuzzing assume sequential, input-determined behavior. An entire kernel
sub-field exists just to control interleavings (\Cref{sec:discovery}).

\myparagraph{C2: Implicit, unspecified invariants.} Kernel correctness rests on
conventions no machine-checkable artifact records: lock-ordering discipline,
reference-count balance, RCU grace periods, the ban on sleeping in atomic
context, object-ownership rules. No test suite encodes them. This is the deepest
difference from user-space APR, whose generate-and-validate loop relies on tests
as a proxy for the specification.

\myparagraph{C3: Cross-syscall, long-lived state.} Kernel objects persist across
system calls, so triggering a bug requires a precise sequence that drives the
kernel into a particular state, and the observable symptom can be far removed
from the offending instruction. User-space targets are frequently single-input,
their crashes closer to their causes.

\myparagraph{C4: Hardware and peripheral dependence.} Device drivers constitute
the majority of kernel code and depend on physical devices, memory-mapped I/O,
DMA, interrupts, and firmware. Exercising or fixing them may require hardware
\edited{unavailable in many test environments}.

\myparagraph{C5: No fault isolation, whole-system blast radius.} The kernel has
no process boundary to contain a fault: a single bug can corrupt arbitrary system
state, failures can be silent, and a benign-looking \texttt{WARNING} may conceal
an arbitrary write. A user-space crash is contained and cheap to roll back; a
faulty kernel patch can render the system unbootable.

\myparagraph{C6: Architecture/configuration multiplicity and multi-tree
deployment.} One kernel source compiles to many architectures and thousands of
configuration options, and ships through mainline plus numerous stable and vendor
trees. A fix must hold across that space and be propagated to every affected
tree, an entire class of work (backporting, patch-presence testing) with no
user-space analog.
\subsection{The Challenges in the Wild}
\label{sec:ch-wild}

The six properties above are not abstractions. We mined the syzbot-fixed corpus
of \Cref{sec:measurement} for each property's footprint, tagging every fix by
lexical and structural signals in its diff, commit message, and review
threads,\footnote{Percentages in this subsection are computed over the same
6{,}946-bug corpus of \Cref{sec:measurement}; the per-challenge taggers are
released with our artifact.} and present one representative merged patch per
challenge; for space, we include in-paper code examples only for C2, C3, and
C6.

\myparagraph{C1, a lock-ordering fix in \texttt{io\_uring}.}
The normal I/O path acquires \texttt{uring\_lock} then the \texttt{seq\_file}
lock; the \texttt{/proc} fdinfo path acquires them in the opposite order. The fix
must reason about the global lock order rather than any single path, breaking the
cycle with a \texttt{trylock}. In our corpus, 10.9\% of all fixes edit a locking or memory-ordering
primitive, and concurrency-class bugs lack any reproducer 44.7\% of the time
versus 22.4\% for the rest.

\begin{figure}[!htb]
  \centering
  \includegraphics[width=0.92\columnwidth]{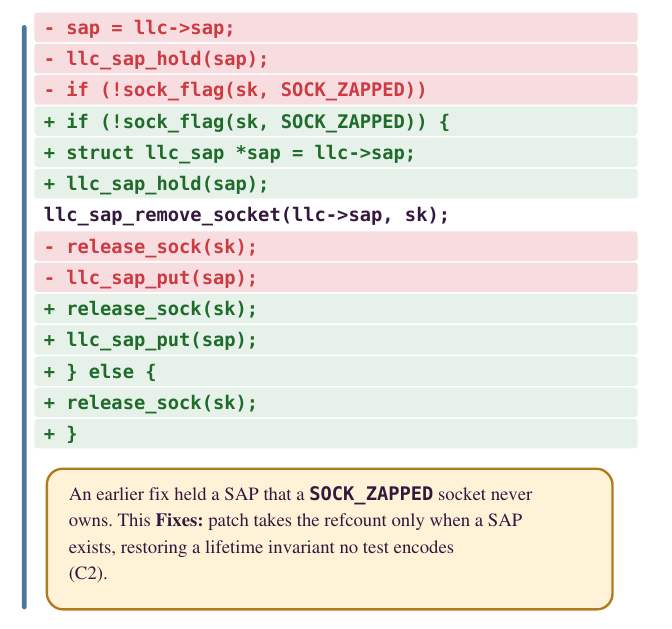}
  \caption{C2 in the wild: a reference-count fix in \texttt{net/llc} that repairs an earlier fix.}
  \label{fig:code-c2}
\end{figure}

\myparagraph{C2, a reference-count fix that repairs another fix.~(\Cref{fig:code-c2})}
An earlier syzbot fix added an unconditional
\texttt{llc\_sap\_hold}/\texttt{put} pair to keep a SAP alive across
\texttt{release\_sock()}, thereby violating a different implicit invariant: a
\texttt{SOCK\_ZAPPED} socket has no SAP at all. The follow-up patch (tagged
\texttt{Fixes:} the first one) restores an object-lifetime rule no test suite
encodes.

Such fix-of-a-fix chains are measurable: at least 4.1\% of corpus fixes repair
another fix in the corpus, and at least 2.1\% were themselves later repaired
again, a direct lower bound on invariant-violating ``complete'' patches.

\begin{figure}[!htb]
  \centering
  \includegraphics[width=0.92\columnwidth]{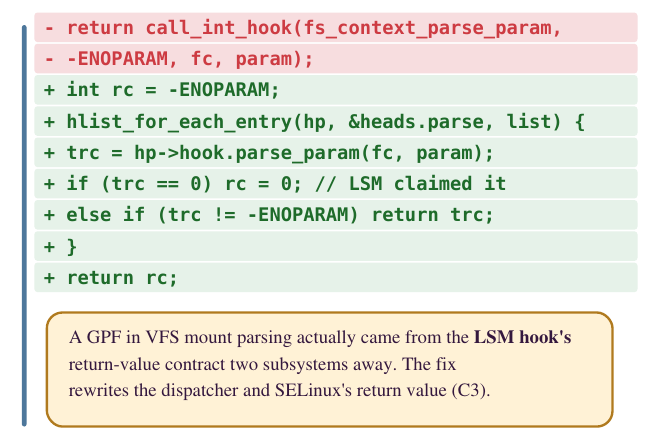}
  \caption{C3 in the wild: a crash far from its cause, fixed in \texttt{security/security.c}.}
  \label{fig:code-c3}
\end{figure}

\myparagraph{C3, a crash far from its cause.~(\Cref{fig:code-c3})}
A general-protection fault manifested in VFS mount-parameter parsing, but the
defect lived in the LSM layer, where stacked security modules disagreed about a
hook's return-value contract. The fix rewrites the hook dispatcher in
\texttt{security/security.c}, two subsystems away from the crash site.

In the corpus, 10.3\% of fixes land outside the crashing directory and
4.2\% land in a different subsystem entirely, and those displaced bugs take a
median 55 days to fix versus 33 for the rest.

\myparagraph{C4, a driver fix validated only by emulation.}
A managed-buffer leak in ALSA PCM hid on the release path that calls the
driver's \texttt{hw\_free} callback directly. The fix factors the callback
handling into one helper invoked from both paths. The defect lives behind a
device-operations interface, and like the 16.1\% of corpus fixes that touch a
driver or sound path, its validation rests on syzbot's emulated devices rather
than the physical hardware it abstracts.

\myparagraph{C5, a benign warning concealing a bounds bug.}
syzbot reported only a \texttt{WARNING} in netlink's extended-ack path.
The fix reveals the substance, as the bounds check for the attribute pointers
compared against the wrong buffer, so the offset written back to user space
could be computed from an address outside the message payload.
About a quarter (26.5\%) of corpus reports carry a benign-looking symptom class
(\texttt{WARNING}, hang, stall), and for 10.8\% of those the merged fix edits
memory-safety-relevant code, the SyzScope risk-inversion at corpus scale.

\begin{figure}[!htb]
  \centering
  \includegraphics[width=0.92\columnwidth]{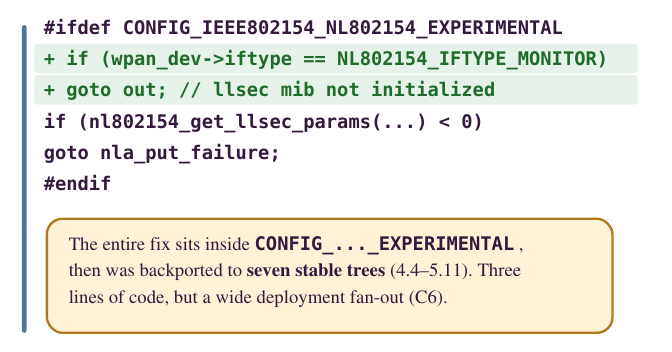}
  \caption{C6 in the wild: a config-conditional ieee802154 fix backported to seven stable trees.}
  \label{fig:code-c6}
\end{figure}

\myparagraph{C6, a config-conditional fix that shipped to seven trees.~(\Cref{fig:code-c6})}
An ieee802154 crash existed only under
\texttt{CONFIG\_IEEE802154\_NL802154\_EXPERIMENTAL}, and the entire fix sits
inside that guard; the patch was then carried into seven stable trees (4.4
through 5.11). The fix itself is three lines, and the C6 burden is the deployment
fan-out around it. Tagging only on unambiguous evidence (explicit \texttt{Cc:\
stable}, config-conditional code, or an \texttt{arch/} file), 17.2\% of corpus
fixes carry a C6 footprint, a lower bound: a further 2{,}329 fixes appear in
stable backport threads (median four trees each) without an explicit tag.

\subsection{The Burden Is Asymmetric}
\label{sec:ch-asym}

\Cref{fig:classification} regroups the surveyed systems along both axes at once,
the lifecycle stage each addresses and the challenge it confronts. The challenges
concentrate toward the back of the pipeline, and the kind of difficulty differs
by stage. At discovery, even high-burden properties are triggering difficulties
that search can amortize away (perturbing schedules for C1, emulating devices for
C4, inferring syscall dependencies for C3): one only needs to provoke the
property once, and continuous fuzzing has unbounded attempts. The later stages
face reasoning difficulties that search cannot dissolve: preserving the global
locking discipline (C1), respecting invariants written down nowhere (C2),
ruling out sibling instances across all configurations (C2, C6), or confirming a
driver fix without the device (C4), while a wrong answer can corrupt the whole
system (C5). These tasks lack exactly what would make them automatable, a
specification, a test oracle, executable hardware, the safety of isolation;
user-space APR matured because it has all four. This asymmetry is, we argue, the
structural reason the gradient exists, and our measurement
(\Cref{sec:measurement}) shows the cost is highest for exactly the bug classes
(concurrency, use-after-free) whose challenges (C1, C2) are hardest to reason
about.

\section{S1: Bug Discovery}
\label{sec:discovery}

Discovery is the stage at which a latent defect is exposed as an observable
failure (a crash, sanitizer report, or analyzer warning). It is the most
thoroughly automated stage of the lifecycle and the reason the rest of the
pipeline is under pressure: continuous fuzzing and whole-kernel static analysis
produce candidate bugs faster than downstream stages can absorb them. As
discovery is well served by existing surveys~\cite{fuzzsurvey}, we keep this
section compact, covering dynamic (\Cref{sec:disc-fuzz}) and static
(\Cref{sec:disc-static}) discovery and the learning-augmented turn
(\Cref{sec:disc-llm}); we tabulate representative systems per stage and the
full S1 classification\edited{, including each system's technique families,}
in the appendix.

\subsection{Dynamic Discovery: Kernel Fuzzing}
\label{sec:disc-fuzz}

Coverage-guided fuzzing is the dominant kernel bug-finding technique, anchored in
practice by \textsc{syzkaller} and its continuous-integration front end
syzbot~\cite{syzkaller,syzbot}. A kernel fuzzer must solve three problems that
distinguish it from user-space fuzzing: obtain coverage feedback from privileged
code, generate structured sequences of interdependent system calls, and reach
deep states guarded by complex preconditions. The literature maps cleanly onto
these problems.

\myparagraph{Coverage feedback and execution.} \textsc{kAFL} established that
hardware-assisted tracing with a thin hypervisor yields general, low-overhead
coverage even for closed-source kernels~\cite{kafl}; a body of follow-on work
drives down execution and instrumentation cost through emulation, VM
checkpointing, snapshotting, binary-only sanitization, and richer feedback
signals~\cite{unicorefuzz,agamotto,horus,nyx,bokasan,liu2024leveraging}. With
feedback largely commoditized, the field's attention shifted to input structure
and state.

\myparagraph{Syscall structure and dependencies.} Because kernel state is built
across syscall sequences, much of the field improves how sequences are
constructed, by distilling seeds from traced syscall logs (\textsc{MoonShine}),
learning inter-syscall influence and dependency relations, or casting mutation
and scheduling as learning problems~\cite{moonshine,healer,mock,actor,syzvegas,snowplow}.
A complementary line confronts the specification bottleneck: syzkaller's
effectiveness depends on hand-written syscall descriptions, so a series of
systems infers interface models for closed-source kernels or generate
descriptions automatically from the kernel--driver
contract~\cite{imf,syzgen,ksg,syzdescribe,chen2024syzgen++}, while \textsc{FuzzNG}
sidesteps descriptions entirely by reshaping the input space around file
descriptors and user pointers~\cite{bulekov2023no}.

\myparagraph{Drivers and peripherals.} Driver code is vast, hardware-dependent
(challenge~C4), and a disproportionate source of bugs, motivating fuzzers that
decouple drivers from physical devices, by reconstructing ioctl interfaces
(\textsc{DIFUZE}), synthesizing or simulating fake device inputs, and extending
interface-aware fuzzing to macOS and Windows
kernels~\cite{difuze,drfuzz,printfuzz,kextfuzz,ntfuzz}. The USB stack, a large
remote attack surface, is reached by device emulation and replay-guided
fuzzing~\cite{usbfuzz,reusb}.

\myparagraph{File systems and concurrency.} Stateful subsystems need
domain-specific input models: \textsc{JANUS} and \textsc{Hydra} jointly mutate
file-system images and operations~\cite{janus,hydra_sosp}. Concurrency
bugs require controlling interleavings (challenge~C1), not just inputs.
\textsc{Razzer} pairs static race candidates with deterministic
scheduling~\cite{razzer}, and successors explore interleaving segments,
inter-thread communication, data-race fuzzing for file systems, and learned
guidance~\cite{segfuzz,snowboard,krace,snowcat}; statically, \textsc{DCUAF} mines
concurrent use-after-free from lock patterns~\cite{dcuaf}.

\myparagraph{Reaching deep state.} Coverage plateaus because many unreached
branches depend on hard-to-synthesize kernel state~\cite{depchallenge}. Responses
track state variables, add symbolic execution for guarded branches, or follow
reference-count state~\cite{statefuzz,hfl,countdown}. A directed strand focuses
scarce fuzzing budget on suspect code, target sites, and risky recent
changes~\cite{syzdirect,syzrisk,directed_industry}. Several papers document the
continuous setting directly~\cite{enterprise,continuous_notes}, which we revisit
as evidence for the crash-to-patch gap (\Cref{sec:measurement}).

\subsection{Static Discovery: Whole-Kernel Analysis}
\label{sec:disc-static}

Static analysis trades soundness and false positives for the ability to reason
about code paths fuzzing rarely reaches and to target specific bug classes.
\textsc{DR.~CHECKER} pioneered a ``soundy'' driver analysis~\cite{drchecker}, and
\textsc{K-Miner} partitioned whole-kernel analysis per syscall~\cite{kminer}. A
productive line infers implicit security rules from the kernel itself, flagging
missing checks, lacking-recheck bugs, use-before-initialization, and double-fetch
windows~\cite{crix,lrsan,ubitect,doublefetch}; others model a single defect
family, such as object-ownership leaks or custom-allocator memory
corruption~\cite{kmeld,goshawk}. \textsc{Coccinelle} occupies a special place:
its semantic patches both find pattern bugs and fix them at scale, foreshadowing
S3~\cite{coccinelle}. Further lines reach the compiler and binary layers,
catching unstable code discarded under undefined behavior~\cite{optsafe} and
memory bugs in binary-only kernels~\cite{digtool}.

The defining tension is precision against scale. Path-sensitive typestate
analysis, on-demand SMT-checked path constraints, incremental analysis across
revisions, and cross-entry taint chaining all sharpen whole-kernel
precision~\cite{li2022path,liu2021kubo,zhai2022progressive,zhang2021statically}.
Because the imprecise first stage can emit tens of thousands of candidates,
recent work pairs these pipelines with LLMs to prune false
alarms~\cite{li2024enhancing,li2025towards}. A second thread targets bugs that
span entry points and object lifetimes, connecting a free in one syscall to a use
in another, racing device cleanup against concurrent syscalls, and checking
allocation intention, accounting, reference-count consistency, and permission
propagation~\cite{zhang2025statically,ma2023top,liu2024detecting,yang2022making,cred,cid,pex}.

Where no specification is written down, the analysis recovers one, from the
kernel's own security checks, error-handling structure, historical fixes, or even
the Kconfig option space~\cite{lu2019automatically,dossche2024inference,zhong2020inferring,oh2021finding}.
Representation choices matter too, from code property graphs to formalized
double-fetch conditions, binary-level recovery, learned features, and static
reasoning about network side
channels~\cite{yamaguchi2014modeling,xu2018precise,garmany2019static,li2018vuldeepecker,man2025scad}.

\subsection{The Learning-Augmented Turn}
\label{sec:disc-llm}

LLMs first entered the lifecycle at discovery, and most maturely at its
specification bottleneck: \textsc{KernelGPT} synthesizes syscall descriptions
that previously required experts~\cite{kernelgpt}, and \textsc{KNighter}
synthesizes checkers, rather than findings, transferring analyst intent into
reusable analyses~\cite{knighter}. The pattern is telling: at S1, LLMs scale
human expertise into automation, not replace an already automated step.

\begin{takeaway}
Discovery is no longer the dominant bottleneck. It is not ``solved'' (hardware
modeling, semantic bugs, and interleaving control remain open), but coverage
feedback is commoditized, and the live frontiers raise the rate of an
already-overflowing pipeline. LLMs here scale human expertise (specifications,
checkers) into automation rather than displacing it.
\end{takeaway}

\begin{openproblem}
Discovery is optimized in isolation from the downstream pipeline: fuzzers and
analyzers are evaluated on bugs found, not bugs fixed. No discovery technique we
surveyed prioritizes findings by downstream fixability, patch availability, or
maintainer load, even though the binding constraint has moved downstream.
\end{openproblem}

\section{S2: Triage and Understanding}
\label{sec:triage}

Discovery produces raw failures, and triage turns a failure into something a
developer can act on: a syzbot crash must be deduplicated against known reports,
its impact, severity, and exploitability assessed, its root cause identified,
and, when a fix already exists upstream, its fixing commit located. This is where
the automation gradient first bends. Several triage subtasks are automated, but
the central one, root-cause analysis, remains expert-driven and is the practical
throttle on everything downstream; we tabulate representative systems for
this stage in the appendix (\Cref{app:survey}).

\myparagraph{Deduplication.}
At syzbot scale, the same defect surfaces under many distinct crash signatures,
inflating the apparent bug count. Mu et al.~\cite{dupreports} performed the
defining study of duplicated kernel bug reports, showing that naive
title/stack-trace bucketing over-merges and under-merges;
\textsc{SyzRetrospector} attacks the same identity problem from the provenance
side~\cite{syzretrospector}, and \textsc{Igor} clusters crashes on root cause
rather than surface signature~\cite{igor}. Deduplication is automated but
imperfect, and its errors propagate: a mis-merged report hides a distinct bug,
while an over-split one wastes triage effort.

\myparagraph{Impact and severity.}
Not all crashes deserve equal attention, and a fuzzer's reported symptom often
understates the true risk. \textsc{SyzScope}~\cite{syzscope} showed that a large
fraction of bugs syzbot labels ``low-risk'' in fact harbor high-risk primitives
such as control-flow hijack, by symbolically exploring the states reachable from
the crash. Related work recomputes severity per derived kernel version
(\textsc{DiffCVSS}), applies LLMs to streamline CVE/CVSS labeling, and predicts
where risk concentrates with metric and text-mining
models~\cite{diffcvss,llmtriage,vulnpredict}.

\myparagraph{Exploitability as a triage signal.}
A bug's exploitation potential is a strong prioritization signal, and a line of
work estimates it automatically, by exploring the alternative error behaviors a
bug can manifest, extracting the capabilities of out-of-bounds writes,
reasoning about leak chains and use-after-free exploitation, evaluating
control-flow-hijack primitives, and testing whether an upstream proof-of-concept
fires on the downstream distributions that actually ship the
code~\cite{grebe,koobe,kleak,fuze,kepler,syzbridge}. We include these works as
triage signals (they answer ``does this bug matter?'') and deliberately exclude
the orthogonal concern of building deployable exploits or runtime defenses,
which a companion SoK covers~\cite{sokhardening}.

\myparagraph{Root-cause analysis, the throttle.}
\edited{Bridging the gap between a crash symptom and its root cause remains
the hardest triage problem (C3). Prior tools operationalize root cause in
three distinct categories: (i) triggering conditions that activate the bug
(e.g., \textsc{AURORA}~\cite{aurora}); (ii) faulty instructions that pinpoint
the defective code statement (e.g., \textsc{ARCUS}~\cite{arcus}); and (iii)
vulnerability-introducing commits that identify the historical change for
regression tracking. Today, tools in all three categories remain heavyweight
and offline, while direct LLM reasoning over ungrounded traces risks
hallucinated explanations. Consequently, downstream patch
generation (S3) cannot proceed without an actionable cause.}

\myparagraph{Fix localization and patch--bug correlation.}
A related triage task links bugs to patches. Locating the security-relevant
commit for a disclosed vulnerability is itself hard, and a line of work ranks
candidate fixing commits, untangles security-relevant hunks from entangled
commits, and recognizes security patches in source or
binaries~\cite{patchscout,dispatch,spain,spi}. These tasks recur in S4
(was this bug actually fixed?) and S5 (is this patch security-relevant?), making
triage and the later stages mutually dependent.

\begin{takeaway}
Triage is partially automated and partially stuck. Deduplication, impact
re-ranking, severity, and exploitability estimation all have automated solutions,
but root-cause analysis, the prerequisite for any repair, remains heavyweight and
expert-driven at kernel scale. The gradient bends here: the pipeline can rank and
label its backlog automatically, but cannot yet explain it automatically.
\end{takeaway}

\begin{openproblem}
Scalable, continuous root-cause analysis is missing. Existing tools are precise
but per-bug and offline; the syzbot setting needs root-cause explanations at
fuzzing throughput, attached to reports automatically. Whether LLMs combined
with execution traces can close this gap is open.
\end{openproblem}

\section{S3: Patch Generation}
\label{sec:generation}

Patch generation synthesizes a candidate fix for a triaged bug. This is the
stage where the automation gradient is steepest: despite a decade of automated
program repair (APR) in user space, kernel-native repair is nascent, and the few
systems that exist report low yields of accepted patches. We explain why the
kernel is hard for repair (\Cref{sec:gen-why}), then survey the three lines that
exist (\Cref{sec:gen-llm}--\ref{sec:gen-backport}), deferring the rich user-space
APR taxonomy to the AVR SoK~\cite{sokavr} as contrast; representative systems
for this stage are tabulated in the appendix (\Cref{app:survey}).

\subsection{Why the Kernel Resists Automated Repair}
\label{sec:gen-why}
User-space APR assumes a property the kernel violates: a comprehensive test suite
that encodes correctness, against which candidate patches can be validated
cheaply~\cite{sokavr}. The kernel offers, at best, a single crashing reproducer,
and ``correct behavior'' is defined by implicit invariants (locking discipline,
reference-count balance, memory ownership, RCU rules, challenge~C2) spread across
millions of lines and rarely written down. A patch must preserve these invariants
under concurrency and across architectures, and a wrong patch can deadlock or
silently corrupt state rather than fail a test. The test-driven
generate-and-validate loop that powers user-space APR is therefore largely
inapplicable, and kernel repair has waited for techniques that can reason from
context rather than from tests, which is where LLMs enter.

\subsection{LLM Repair Agents}
\label{sec:gen-llm}
The current frontier is agentic LLM repair. \textsc{kGym}/\textsc{kBench}
provided the enabling platform that compiles, boots, and tests kernels at scale,
with a dataset of real syzbot bugs and developer fixes~\cite{kgym}. Built on it,
\textsc{CrashFixer} is the first LLM repair agent targeting the Linux kernel,
mirroring a developer's investigation workflow at the scale of
20M~LOC~\cite{crashfixer}; \textsc{PatchIsland} orchestrates multiple agents in a
continuous-repair pipeline coupled to fuzzing~\cite{patchisland}; and ``beyond
crash-to-patch'' work studies how an initial fix is refined rather than one-shot
generation~\cite{beyondc2p}. Reported accepted-fix rates remain low
(single digits on kBench-style benchmarks~\cite{kgym}), and most evaluations
measure reproducer resolution, not upstream acceptance. In user space, by
contrast, prompt-based agents already fix substantial bug counts
cheaply~\cite{chatgpt162,hitlapr}. The kernel gap is one of validation
infrastructure and invariants, not of generation capability per se.

\subsection{How Far Do Current Methods Get?}
\label{sec:gen-bench}
To measure the generation gap directly, we benchmark \edited{thirteen} LLM patch-generation
\edited{configurations, eleven without and two with a localization oracle,} on 80 crashes sampled from our dataset (\Cref{sec:measurement}). We restrict
to evolution-stage bugs, whose first upstream fix was itself revised, so each is
hard and carries review signal, and we score every candidate patch on two axes:
localization (does it edit the files the developer fix touched?) and repair (a
semantic judge decides whether the patch resolves the same root cause as the
merged fix). \edited{While recent work cautions that LLM-as-a-judge evaluations can introduce label inaccuracy and bias \cite{llmpitfalls}, we mitigate this risk through multi-model cross-validation and systematic human inspection. Two judge models score each candidate independently, each with a written rationale, and two authors re-judge every case on which the models disagree. Across a sampled subset of 100 candidate patches evaluated independently by both authors to verify agreement, author verdicts agree with the judge’s on 89\% of cases and with each other on 94\%.} The methods span one-shot prompting, sampling
(Best-of-N~\cite{selfconsistency}), conversational and self-directed repair
(ChatRepair~\cite{chatgpt162}, ThinkRepair~\cite{thinkrepair}), autonomous agents
(RepairAgent~\cite{repairagent}, AutoCodeRover~\cite{autocoderover},
RGym~\cite{rgym2025}), the CrashFixer pipeline~\cite{crashfixer}, and kGym's
file/function localization oracles~\cite{kgym}, all run on the kGymSuite
platform~\cite{kgymsuite}. \edited{We also test the two most recent code
agents, Claude Fable 5 agent and Codex 5.6 agent, at high reasoning effort.}

\edited{\Cref{tab:patchgen} reports the results. We score localization in three ways across the 80 bugs: whether a patch touches at least one file from the developer fix (any), matches the exact file set (file), or matches the exact function set (func). Structure and macro edits sit outside functions, so we count them at file level. For repair, we record the number of patches the judge rates as FIXED or PARTIAL. Threats to validity are discussed in the appendix. The result is stark. Localization is far easier than repair, but not solved. Methods edit at least one correct file 55-66\% of the time, and a file or function level oracle pushes this to 95-100\%. Pooling the eleven non-oracle configurations over all 80 bugs, only 43.5\% of candidate patches recover the exact file set and 14.1\% the exact function set, and on the 18 multi-file and 31 multi-function fixes no method recovers the complete set, failing by omission rather than by editing irrelevant locations.} 

End-to-end repair never exceeds 5/80 (6\%), and the oracle rows make the point sharpest. Told exactly which file to change, models still fix 0/80, and told the exact function, only 2/80. The wall is synthesizing a correct fix, not finding where it goes. A weaker GPT-4o-mini base repaired essentially nothing (0--1/80); only with a stronger base and real code search do the better designs begin to register. One honest caveat is that real compile-and-reproduce feedback, the engine of several of these methods, was out of reach at this scale, \edited{so the feedback-driven rows may underestimate those methods. While such
feedback may improve these results, the 6\% repair rate shows that generating
correct kernel logic remains the central bottleneck, though it is not a
ceiling on capability.}

\begin{table}[t]
  \centering
  \caption{End-to-end LLM patch generation on 80 evolution-stage syzbot crashes.}
  \label{tab:patchgen}
  \scriptsize
  \renewcommand{\arraystretch}{1.05}
  \setlength{\tabcolsep}{4pt}
  \begin{tabular}{@{}l ccc cc@{}}
    \toprule
    & \multicolumn{3}{c}{\textbf{Local.\,\%}} & & \\
    \cmidrule(lr){2-4}
    \textbf{Method} & \textbf{any} & \edited{\textbf{file}} & \edited{\textbf{func}} & \textbf{Fixed/80} & \textbf{Part./80} \\
    \midrule
    one-shot baseline                          &  65 & \edited{48} & \edited{15} & 2 & 21 \\
    Best-of-N~\cite{selfconsistency}           &  55 & \edited{44} & \edited{12} & 3 & 20 \\
    ChatRepair~\cite{chatgpt162}               &  55 & \edited{39} & \edited{12} & 0 & 6 \\
    ThinkRepair~\cite{thinkrepair}             &  66 & \edited{45} & \edited{15} & 0 & 23 \\
    RepairAgent~\cite{repairagent}             &  59 & \edited{41} & \edited{15} & 1 & 4 \\
    AutoCodeRover~\cite{autocoderover}         &  55 & \edited{41} & \edited{14} & 2 & 9 \\
    {RGym} SimpleAgent~\cite{rgym2025}         &  60 & \edited{42} & \edited{14} & 2 & 14 \\
    {RGym} ExplorationAgent~\cite{rgym2025}    &  56 & \edited{44} & \edited{14} & 4 & 16 \\
    CrashFixer~\cite{crashfixer}               &  61 & \edited{45} & \edited{14} & 5 & 20 \\
    \midrule
    \multicolumn{6}{@{}l}{\footnotesize\itshape \edited{recent code agents:}}\\
    \edited{Claude Fable 5 agent (high)}       &  \edited{62} & \edited{45} & \edited{15} & \edited{3} & \edited{19} \\
    \edited{Codex 5.6 agent (high)}            &  \edited{64} & \edited{46} & \edited{16} & \edited{4} & \edited{17} \\
    \midrule
    \multicolumn{6}{@{}l}{\footnotesize\itshape given a localization oracle (target files / functions):}\\
    kGym oracle, files~\cite{kgym}             & 100 & \edited{95} & \edited{21} & 0 & 10 \\
    kGym oracle, +functions~\cite{kgym}        &  95 & \edited{80} & \edited{64} & 2 & 4 \\
    \bottomrule
  \end{tabular}
\end{table}

\begin{finding}
On 80 hard kernel crashes, LLM patch generation \edited{hits at least one correct
file 55-66\% of the time, and 95-100\% given an oracle,} but repairs at
most 6\%. Even told the exact file
and function to edit, models fix $\le$2/80. Kernel repair is bottlenecked on
synthesizing a correct fix under implicit invariants, not on locating it.
\end{finding}

\subsection{Semantic Transformation}
\label{sec:gen-transform}
Predating LLMs, the kernel community automated repair through semantic patches.
\textsc{Coccinelle}'s SmPL lets a maintainer express a cross-tree change as a
near-patch and apply it everywhere; over a decade it is credited with thousands
of commits, making it the most successful deployed kernel repair technology by
volume, with roots in automating collateral evolutions as driver APIs
change~\cite{coccinelle,collateral}. These approaches are fully automated once a
human writes the rule: they fix known patterns at scale rather than synthesizing
novel fixes.

\subsection{Backporting, the Mature Kernel Repair Task}
\label{sec:gen-backport}
The one kernel repair task with robust automation is backporting, which carries a
mainline fix into older stable trees where names, locations, and surrounding
logic differ (challenge~C6). \textsc{FixMorph} synthesizes a transformation rule
from a mainline patch and applies it to the older version, correctly backporting
75\% of 350 patches~\cite{fixmorph}; companions classify which commits are
stable-worthy, resolve conflicts against divergent downstream code, and quantify
how much porting still falls to humans~\cite{patchnet,patchscope,patchporting}.
Backporting is tractable precisely because it has an oracle the rest of S3
lacks: the original patch already encodes the correct fix, so the task is
transfer rather than synthesis.

\begin{takeaway}
Kernel-native patch generation is the least mature stage. The test-driven loop
behind user-space APR does not transfer, because kernel correctness lives in
implicit invariants rather than test suites. The only mature repair tasks are
those with a built-in oracle, pattern fixing (\textsc{Coccinelle}) and
backporting (\textsc{FixMorph}), where a human or an existing patch supplies the
specification. LLM agents are the frontier for novel fixes but report low
accepted-patch yields.
\end{takeaway}

\begin{openproblem}
Repair without a test oracle. The central open problem is generating kernel
patches that provably preserve invariants absent comprehensive tests. This needs
(i) machine-checkable encodings of kernel invariants (locking, refcount, RCU,
ownership) usable as repair constraints, and (ii) benchmarks that score
upstream-accepted fixes, not just reproducer resolution. Today's agents optimize
the latter while ignoring the former.
\end{openproblem}

\section{S4: Patch Validation}
\label{sec:validation}

A candidate patch, whether written by a developer or generated by an agent, is
not a fix until it is shown to resolve the bug, preserve functionality, and
leave no residual or newly introduced defect. Kernel validation inherits the
oracle problem that hampers generation (S3): without comprehensive tests,
``correct'' is hard to establish mechanically, and ``complete'' is harder
still.

\myparagraph{Correctness checking.}
The most developed validation task asks whether a patch is correct.
\textsc{KLAUS} attacks this directly for the kernel: from a study of 182
incorrectly developed patches, it observes that errors usually stem from the
patch's altered read/write operations, and steers a fuzzer toward the affected
contexts, confirming and fixing 25 incorrect patches upstream~\cite{klaus}. In
user space, correctness assessment has a longer history~\cite{patchimpact}, and
LLM-as-judge schemes with a human in the loop have recently been proposed to
scale it~\cite{llmjudge}. These reduce, but do not eliminate, the manual burden.

\myparagraph{Completeness and incomplete fixes.}
Correctness is necessary but not sufficient: a patch can resolve the reported
crash yet leave sibling instances unfixed, or introduce a new defect. Incomplete
fixes are common enough in the kernel to be a named, studied phenomenon. Liu et
al.~\cite{incompletefix} identify three recurring root causes (developers misled
by the surface symptom, neglecting similar modules, or introducing a new semantic
error) and build a similarity-based detector that uncovered previously unknown
cases. This is precisely the failure mode automated generation (S3) is most
prone to, and it is barely tooled: detecting that a fix is complete has no
scalable, deployed solution.

\myparagraph{Patch presence testing.}
Validation also has a downstream-deployment dimension: given the fragmented
ecosystem of vendor and distribution kernels (challenge~C6), is a particular tree
actually patched? \textsc{PDiff} decides whether a known fix is present despite
version drift~\cite{pdiff}, and \textsc{PS3} sharpens this to a precise test from
a semantic signature of the patch~\cite{ps3}. The same version-alignment
reasoning that makes backporting hard (S3) makes verifying deployment hard here.

\myparagraph{Benchmarks for validation.}
Recent work argues that validation itself needs better ground truth, formalizing
patch validation around a proof-of-concept plus functional and synthesized unit
tests~\cite{pvbench}. As with generation, the scarcity of kernel benchmarks with
reproducers, fixes, and completeness oracles is a limiting factor we return to in
\Cref{sec:measurement}.

\begin{takeaway}
Validation is where the oracle problem bites hardest. Correctness has partial,
kernel-specific automation (\textsc{KLAUS}) and patch-presence testing is solved
(\textsc{PDiff}), but completeness, did the patch fix all instances and introduce
none, is essentially manual, even though it is the dominant failure mode of
automatically generated patches.
\end{takeaway}

\begin{openproblem}
Automated completeness checking. We lack scalable methods to decide whether a
kernel patch fixes every sibling manifestation of a defect and introduces no
regression. As LLM agents (S3) generate more patches, the validation bottleneck,
not the generation bottleneck, will dominate. Co-designing generation with
completeness-aware validation (e.g., generating the sibling-instance test
alongside the patch) is unexplored.
\end{openproblem}

\section{S5: Community Integration}
\label{sec:integration}

A validated patch is still not a deployed fix: it must be posted, reviewed,
revised, accepted by a maintainer, merged, and backported to the stable trees
real systems run, and even then deployment may demand a reboot that live kernel
updating tries to avoid~\cite{seamless}. This is the automation gradient's
floor: integration is governed not by an algorithm but by a socio-technical
process of mailing-list review, maintainer attention, and human judgment, the
stage the security literature has most neglected even as it has become the
binding constraint.

\myparagraph{Code review.}
Review is the gate every kernel patch passes through, and it is overwhelmingly
manual. One line of work automates parts of it, learning the contributor and
reviewer sides, pre-training on code-change/review data, and generating review
comments, with the LLM era adding benchmarks, developer-aligned feedback, and
workflow studies~\cite{tufano2021,codereviewer,auger,llmreviewbench,dpofplus,llmreviewstudy}.
Crucially, almost all of this work is evaluated on general open-source corpora,
not the kernel, whose review norms (LKML etiquette, Signed-off-by chains,
subsystem trees) differ sharply. A second line studies review as human practice,
characterizing its expectations and outcomes, its effect on quality, reviewer
participation, and review strategies~\cite{bacchelli2013,mcintosh2016,ruangwan2019,goncalves2022};
for the kernel specifically, work documents patch-submission
communication~\cite{tan2019communicate} and shows that incivility on LKML
correlates with rejected changes~\cite{ferreira2021}. Integration outcomes hinge
on human and social factors that no current automation models.

\myparagraph{The maintainer bottleneck.}
The kernel's integration capacity is fundamentally a function of maintainer
bandwidth, and it does not scale with the inflow of patches and bug reports. Zhou
et al.~\cite{zhou2017} show that maintainer workload is highly unbalanced and
that adding co-maintainers yields only sublinear gains, and Tan et
al.~\cite{tan2020} analyze the multiple-committer model's pressure--latency--quality
trade-offs. The strain is visible at the margins, in newcomer
onboarding, reviewer scarcity for Rust-for-Linux, and contributor
retention~\cite{onboarding,rust4linux,womenoss}. After the kernel became a CVE
numbering authority in 2024, CVE volume rose by an order of magnitude, sharply
increasing patching demand~\cite{patchmeifyoucan}. This is the human face of the
crash-to-patch gap.

\myparagraph{Acceptance and disclosure.}
Whether and how fast a patch is accepted has been studied empirically. Jiang et
al.~\cite{jiang2013} find that only a fraction of submitted patches reach a
release and that author experience strongly predicts acceptance speed;
coordination across CVE numbering authorities and disclosure management add
further process latency~\cite{cnacoord,disclosure}; drivers dominate both
regression frequency and fix slowness~\cite{fastfixes}; and static-analysis
alerts are often left unaddressed~\cite{coverityalerts}. These findings quantify,
from the process side, the same delay our measurement (\Cref{sec:measurement})
observes from the data side.

\myparagraph{Datasets that measure the pipeline.}
Finally, integration is where end-to-end datasets live. A multi-level patchwork
dataset links patches, reviewers, and commits across nine years of
LKML~\cite{patchwork}, and for the repair-centric pipeline,
\textsc{kGym}/\textsc{kBench}~\cite{kgym} and live crash-resolution
benchmarks~\cite{livebench} pair reproducers with developer fixes. We use these,
together with the public syzbot dashboard, as the basis for our measurement.

\begin{takeaway}
Integration is the automation-gradient floor. The decisive resources are human:
maintainer attention, review latency, author reputation, even discourse civility.
Automation here is nascent and, tellingly, almost never kernel-specific; the
richest body of evidence is descriptive (empirical SE studies), not prescriptive.
\end{takeaway}

\begin{openproblem}
Closing the loop, not just generating patches. The field optimizes generation
while the binding constraint is integration: kernel-aware review assistance,
maintainer-load-aware routing, and agents that carry a fix through revision
rounds are all missing. Until automation targets integration, more generated
patches may worsen, not relieve, the maintainer bottleneck.
\end{openproblem}

\section{Measuring the Crash-to-Patch Gap}
\label{sec:measurement}

The preceding sections argue qualitatively that automation thins toward the back
of the pipeline. We now ground that claim by measuring the crash-to-patch gap on
bugs that traverse the entire lifecycle, decomposing it to locate where the time
is actually spent.

\subsection{Dataset and Method}
We assembled a dataset of 6{,}946 Linux kernel bugs that syzbot reports as fixed,
each linked through its full lifecycle: first and last crash timestamps, fix
timestamp and merged commit, the reconstructed patch series
(v1$\rightarrow$v2$\rightarrow$\dots) and reviewer threads from
\texttt{lore.kernel.org}, and reproducer availability.\footnote{Collected from
the public syzbot dashboard and kernel git/mail archives; scraper and analysis
scripts are released with the artifact.} When reconstructing review threads we
discard stable-backport batch series and pull-request digests
(\texttt{[PATCH 4.14 000/164]}, \texttt{[GIT PULL]}), which the archive
over-associates with a bug and which would otherwise inflate per-bug discussion
counts by orders of magnitude. Bug-class and subsystem labels were derived by two
authors from report titles and merged-patch paths using a fixed rule set, with
disagreements resolved by discussion.

\edited{We use syzbot because it provides unusually complete public linkage
among crash reports, available reproducers, and fixes. Our results
characterize eventually fixed, syzbot-reported bugs and may not generalize
to out-of-band reports, especially those submitted with patches.} We study the fixed population deliberately: these bugs have a well-defined
crash-to-patch latency, and slow recent bugs are right-censored, so our
latencies are a conservative lower bound on the gap. The snapshot also contains
364 still-open reports, whose open rate is highest for the bug classes hardest
to reason about (\Cref{sec:meas-readiness}), reinforcing the same bias.

\subsection{Crash-to-Patch Latency}
\begin{figure}[t]
  \centering
  \includegraphics[width=\columnwidth]{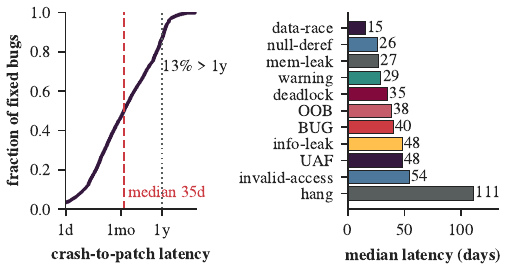}
  \caption{\textbf{Crash-to-patch latency over 6{,}946 fixed kernel bugs. Left:
  CDF on log-time. The median is 35 days but the tail is long (13\% exceed one
  year). Right: median latency by bug class. Semantically diffuse bugs
  (hangs, UAF) linger, while sharp-signature bugs (races, null-deref) close
  fast.}}
  \label{fig:latency}
\end{figure}

The central measurement is the time from a bug's first observed crash to its fix.
The distribution (\Cref{fig:latency}, left) is severe and heavy-tailed: the median
fixed bug takes 35 days to patch, the mean 138; more than half take over a month,
13\% take more than a year, and the slowest waited 7.5 years. For a stage that
produces bugs in seconds of fuzzing, a median month-plus to patch is the
automation gradient made concrete.

The latency also varies by bug class in a telling way (\Cref{fig:latency},
right). Semantically diffuse failures take longest, hangs/stalls (median 111
days) and corrupted-state failures (57), whose symptom sits far from its cause;
classes with a sharp, local signature close fastest, data races (15) and
null-pointer dereferences (26). The bugs hardest to understand (S2) and repair
correctly (S3) are exactly the ones that linger, consistent with our claim that
the back-end stages, not discovery, set the pace.

\begin{finding}
Even among bugs that are eventually fixed, the crash-to-patch gap is large and
heavy-tailed (median 35 days, mean 138, 13\% over a year), and it is a
conservative lower bound: unfixed and slow recent bugs are excluded.
\end{finding}

\myparagraph{The gap is structural, not transient.} One might expect a decade of
improving tooling to have shrunk the gap. The temporal trend says otherwise. The very high medians of 2017--2018
(471 and 291 days) reflect syzbot's launch clearing a backlog of long-latent
bugs; once the pipeline reached steady state in 2019, the median plateaued at
roughly three weeks (15--25 days) and has stayed there for six consecutive years,
even as discovery throughput and LLM tooling advanced. (The dip in the most
recent years is right-censoring, which makes the plateau, if anything,
optimistic.) Better finding has not translated into faster closing.

\begin{finding}
Since the continuous-fuzzing pipeline matured in 2019, median crash-to-patch
latency has held at roughly three weeks for six years. The gap is a stable
structural property of the back end, not a transient that better bug finding
will erode.
\end{finding}

\subsection{Where the Time Goes}
\label{sec:meas-decomp}
\begin{figure}[!htb]
  \centering
  \includegraphics[width=\columnwidth]{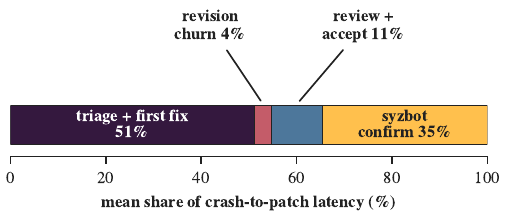}
  \caption{\textbf{Decomposition of crash-to-patch latency into ordered segments, as the
  mean per-bug share of total latency ($n{=}3{,}371$). Just over half the wait
  precedes the first posted patch (triage and first-fix authoring). Explicit
  revision churn is a thin 4\% mean with a long tail, and a third is syzbot's
  post-merge confirmation lag rather than engineering time.}}
  \label{fig:decomp}
\end{figure}

A single latency number cannot say which stage is slow. We therefore decompose
each bug's lifecycle into ordered segments (first crash $\rightarrow$ first patch
posted $\rightarrow$ final patch version $\rightarrow$ merged commit
$\rightarrow$ syzbot marks fixed) using mail and git timestamps, and compute each
segment's share of that bug's total latency (\Cref{fig:decomp}, $n{=}3{,}371$
bugs with a complete, monotone chain).

First, the largest share of the wait, 51\% on average, elapses before the first
patch is even posted: the bug sits after discovery, waiting to be triaged,
root-caused, and turned into a candidate fix. The median time to the first human
reply is 6 days and to the first posted patch 7, though a heavy tail languishes
for months. This is the human attention/throughput bottleneck the gradient
predicts, and it dwarfs the revision loop.

Second, explicit revision churn accounts for only 4\% of total latency on
average, because most accepted fixes are merged at their first or second
version. Revision is nonetheless the failure mode of the hard cases (21\% of
fixes with a reconstructable series needed two or more versions, up to nine), a
thin median with a long tail. The remaining 11\%, time between first posted
patch and merge not explained by visible revisions, is review and acceptance
latency, the patch waiting on a maintainer rather than on its author.

Third, a substantial 35\% of the nominal latency is post-merge: the lag between
the fix landing and syzbot confirming the bug no longer reproduces (median 7
days, mean 71). This is infrastructure latency, not engineering effort. The
engineering gap (crash$\rightarrow$merge) is therefore somewhat shorter than the
headline, with the remaining delay concentrated precisely in the human-bound
front of the back end, in getting a correct first patch written and landed.

\begin{finding}
\edited{About 51\% of measured latency occurs before the first patch, 4\%
during visible revision churn, 11\% during review and acceptance, and 35\%
during syzbot's post-merge confirmation. These timestamps locate the delay but
do not identify a single cause. Bug complexity, subsystem characteristics,
reproducer quality, and maintainer availability may all contribute.}
\end{finding}

\subsection{Heterogeneity Across Subsystems}
\label{sec:meas-subsys}
The gap is also unevenly distributed across the kernel. Among the fifteen
subsystems with the most fixed bugs, median latency spans more
than an order of magnitude. Here \texttt{kernel/bpf} is the slowest by a wide margin
(median 212 days), followed by \texttt{arch/x86} (91),
\texttt{fs/ext4} (85), and cross-cutting \texttt{include/} header changes (77),
while \texttt{fs/io\_uring.c} (6), \texttt{net/sched} (9), \texttt{mm} (17), and
\texttt{net/core} (18) close
fastest. The slow subsystems are those where a fix must satisfy an unusually
demanding correctness bar (the BPF verifier's safety contract) or coordinate
across many drivers, whereas the fast ones tend to have a single responsive maintainer or
a self-contained fix.

Reproducer availability is likewise subsystem-dependent: \texttt{net/core},
\texttt{kernel/bpf}, and \texttt{net/ipv4} bugs lack any reproducer 31-32\% of
the time, whereas device subsystems with concrete trigger paths
(\texttt{drivers/media} 7\%, \texttt{drivers/usb} 11\%) are far better
supplied. A repair agent's applicability is thus gated subsystem-by-subsystem by
the very artifacts the front end does or does not emit.

\begin{finding}
Latency varies by more than 10$\times$ across subsystems (median 6-212 days),
and reproducer scarcity ranges from 7\% to 32\%. Both the delay and the
artifacts automation depends on are kernel-region-specific, and no single
intervention closes the gap everywhere.
\end{finding}

\myparagraph{\edited{Challenge prevalence.}} \edited{Tagging each fixed
bug shows that concurrency (C1) affects 19.3\% of reports, configuration
differences across trees (C6) 17.2\%, and hardware dependence (C4) 16.1\%,
followed by implicit invariants (C2, 8.4\%), cross-syscall state (C3, 7.4\%),
and lack of fault isolation (C5, 2.9\%).}

\myparagraph{\edited{Where the pre-merge delay sits.}} \edited{
Two direct signals place the engineering delay in the human-in-the-loop
stages, where patches are revised and discussed over multiple rounds. Of the 5{,}252 bugs with a reconstructable
patch series, 21\% required two or more revisions before acceptance, the
correctness and completeness gap of S3--S4 manifesting as resubmission, and
discussion is heavy-tailed (median 6 non-bot messages, mean 18, top decile
40). Both signals sit in validation and integration, not in generating a first
candidate. The remaining 35\% of nominal latency is syzbot's post-merge confirmation
lag (\Cref{sec:meas-decomp}). Thus, our
observation covers only the crash-to-merge window, where bug complexity,
subsystem characteristics, and reproducer quality collectively shape the delay
(\Cref{sec:meas-subsys}). We
therefore read the data as locating the delay, not as isolating a single
cause.}

\begin{finding}
\edited{The pre-merge crash-to-patch delay concentrates in the
human-in-the-loop downstream stages.} Some
21\% of fixes take $\geq$2 review-driven revisions and a long tail of bugs draws
40+ reviewer messages, while the median bug spends most of its life simply
waiting for a first fix. \edited{A further 35\% of nominal latency is pipeline
mechanism, syzbot's post-merge confirmation. This is
the pattern the automation gradient predicts.}
\end{finding}

\subsection{Automation Readiness: Reproducers and Patch Shape}
\label{sec:meas-readiness}
Downstream automation (triage, repair agents, validation) depends on a reproducer
to ground its reasoning and check its output, and is easiest when the required
fix is small and local. Yet 26.7\% of even the fixed bugs had no reproducer at
all, and a full 33\% lacked a C reproducer: a quarter of the very bugs humans did
close would have been out of reach for today's reproducer-driven repair agents
(S3) before any modeling limitation applies. Scarcity also tracks bug class: data
races essentially never ship a reproducer (sequential replay cannot capture the
interleaving), and use-after-free and hangs lack a C reproducer 40\% of the time,
exactly the classes whose latency is highest.

The shape of the accepted fix is more encouraging: the median merged fix touches
1 file (57\% single-file) and changes 5 lines (67\% change $\le$10), the regime
where automated repair is most plausible, so the binding constraint is the input
artifacts more than the size of the edit. To make ``repair-readiness'' concrete,
we score each fixed bug against the artifacts current agentic pipelines assume:
a C reproducer, a single-file fix, a small ($\le$50-line) diff, and acceptance
without revision. Only 34\% of fixed bugs satisfy all four, and requiring light
review ($\le$5 messages) drops the share to 18\%; the rest fall outside the
operating envelope today's benchmarks reward.

\begin{finding}
A quarter (26.7\%) of fixed bugs lack any reproducer and a third lack a C
reproducer, yet the typical fix is small (median 1 file, 5 lines). Only 34\% of
fixed bugs are ``repair-ready'' (C reproducer + single-file + $\le$50-line + no
revision), 18\% once light review is also required. The input artifacts, not
the edit size, gate back-end automation.
\end{finding}

\edited{We also map the surveyed techniques onto a grid of bug class by
lifecycle stage to see where dedicated automation exists. The matrix and its
discussion are in the appendix.}

\begin{finding}
No kernel bug class is served by dedicated automation across all four technical
stages. Coverage is dense at discovery and triage and collapses at generation and
validation. The matrix's empty lower-right region is the research frontier this
SoK identifies.
\end{finding}

\section{Discussion and Future Directions}
\label{sec:discussion}

Our survey and measurement converge on one structural fact: automation is
concentrated at discovery and drains away toward integration, and the cost of
that asymmetry, the crash-to-patch gap, is dominated by the stages the security
community has invested in least. We close with cross-cutting directions.

\myparagraph{D1. Rebalance the field from finding to closing.} The marginal
discovered kernel bug is nearly free; the marginal closed bug is expensive and
slow, yet discovery remains the largest category even in our balanced corpus.
Effort should move toward the right half of \Cref{fig:pipeline}, and new
discovery work should be evaluated on its effect on the downstream pipeline.

\myparagraph{D2. Downstream-aware discovery.} Discovery should surface bugs with
the artifacts needed to close them: a quarter of fixed bugs lacked any
reproducer, structurally blocking automated repair. Fuzzers that co-produce a
minimized reproducer, root cause, or fixability estimate would attack the gap at
its source.

\myparagraph{D3. Continuous, grounded root-cause analysis.} \edited{Scaling
root-cause analysis to fuzzing throughput is the primary open challenge in
triage (\Cref{sec:triage}). To prevent ungrounded LLM reasoning and
hallucinations~\cite{llmpitfalls}, models should not guess causes in
isolation. Instead, they should act as hypothesis generators coupled with
dynamic execution feedback (e.g., using lightweight emulation to falsify
candidate predicates). Crucially, downstream repair needs these root causes
formulated as machine-actionable invariants (e.g., locking constraints or
lifetime bounds) rather than human-readable text, supplying the
synthesis specifications currently missing in S3 and S4.}

\myparagraph{D4. Repair and validation without a test oracle, co-designed.} The
steepest part of the gradient (S3--S4) shares one root cause: kernel correctness
lives in implicit invariants, not test suites. Progress requires
machine-checkable encodings of kernel invariants (locking, refcount, RCU,
ownership) usable simultaneously as repair constraints and validation oracles,
and generation that emits sibling-instance and regression tests alongside the
patch; an agent should iterate against such oracles the way a developer
iterates against reviewers.

\myparagraph{D5. Automation aimed at integration.} The gradient's floor (S5) is
where automation is scarcest and least kernel-specific; kernel-aware review
assistance, maintainer-load-aware routing, and agents that shepherd a patch
through revision rounds are all open. We caution that scaling generation without
scaling integration may worsen the bottleneck: machine-generated patches land on
the same finite maintainer attention our data shows is already strained.

\myparagraph{D6. Benchmarks that score the whole lifecycle.} Current kernel
benchmarks~\cite{kgym,livebench} measure reproducer resolution, but almost none
score upstream acceptance, completeness, or invariant preservation, the
properties that actually gate a fix. Benchmarks that reward closing a bug as the
community defines it would realign the field's incentives.

\myparagraph{LLMs as connective tissue.} Across stages, the four LLM roles show
sharply different maturity: artifact generators are effective where they scale
human expertise over a grounded artifact (specifications and checkers at S1,
patches and review comments at S3/S5); classifiers/judges are a useful but
unverified labeling aid (S2, S4); agents are an active but early frontier with
low accepted-fix yields (S3/S5); and reasoning engines for the tasks that lack
an oracle (root cause, invariant-preserving repair, completeness) \edited{remain
the least mature}, because the model cannot self-verify what the kernel never makes
explicit. The productive frontier is the first three roles coupled to oracles,
not the fourth in isolation.

\myparagraph{\edited{A moving snapshot.}} \edited{Recent industry evidence
confirms this gradient. Anthropic's Project Glasswing reported over 10{,}000
vulnerabilities of high or critical severity within a month, but maintainers
have patched only 75 of the 530 bugs reported to them~\cite{glasswing}. The
real challenge has shifted from finding bugs to fixing them. Our automation
levels capture a moving boundary rather than a permanent limit.}

\myparagraph{Relation to prior systematizations.} Kernel-fuzzing
surveys~\cite{fuzzsurvey} organize discovery but not what follows it; the AVR
SoK~\cite{sokavr} systematizes user-space repair, whose central test-suite
assumption fails in the kernel (S3); the kernel-hardening SoK~\cite{sokhardening}
asks how to survive unfixed bugs where we ask how bugs get fixed; and empirical
SE studies~\cite{jiang2013,bacchelli2013,mcintosh2016,kernvulndefenses,opensslempirical,vulnlifetime,refcountsosp}
corroborate our measurement from the process side. \edited{While Alexopoulos et al. ~\cite{vulnlifetime} measure overall vulnerability lifetimes across open-source software, we specifically investigate post-discovery latency and the behavior of automation prerequisites across kernel lifecycle stages.} To our knowledge, this is
the first systematization to span the five stages together and to quantify the
crash-to-patch gap as their unifying consequence.

\section{Conclusion}
\label{sec:conclusion}

We systematized the OS kernel bug lifecycle as a five-stage pipeline governed by
an automation gradient: techniques are mature where bugs are found and grow
sparse and \edited{human-in-the-loop} toward a deployed fix, with LLMs closing the gap only
where coupled to a grounded oracle. Measuring 6{,}946 fixed kernel bugs made the
consequence concrete: a median 35-day wait \edited{that sits in the human-in-the-loop downstream
stages}.
The community has spent a decade learning to find bugs faster than ever; the
next decade's challenge, and this SoK's call, is to learn to close them.

\bibliographystyle{IEEEtran}
\bibliography{refs}

\renewcommand{\bottomfraction}{0.99}
\renewcommand{\textfraction}{0.01}
\setcounter{bottomnumber}{2}
\setcounter{totalnumber}{4}
\setlength{\textfloatsep}{5pt}
\setlength{\floatsep}{4pt}

\appendices
\section{Classification Tables; Open Science and Ethics}
\label{app:survey}
\label{app:discovery}

\myparagraph{A coverage-gap matrix.}
Our final empirical instrument maps the surveyed techniques onto a (bug class
$\times$ lifecycle stage) grid (\Cref{tab:coverage}), marking each cell by the
strongest automation available: a dedicated kernel technique (\cmark), only a
generic one (\pmark), or none (\xmark). The shape mirrors the gradient exactly.
Discovery is uniformly covered; triage has dedicated exploitability tooling for
memory bugs but relies on generic root-cause analysis elsewhere; generation and
validation are mostly \pmark/\xmark, and no bug class enjoys end-to-end dedicated
automation. The empty lower-right of the matrix is the crash-to-patch gap, drawn
at the technique level.

\begin{table}[!htb]
  \centering
  \caption{Coverage-gap matrix: strongest automation per (bug class $\times$
  technical stage). \cmark=dedicated kernel technique, \pmark=generic only,
  \xmark=none; counts are fixed-bug frequencies in our dataset.}
  \label{tab:coverage}
  \scriptsize
  \renewcommand{\arraystretch}{1.05}
  \setlength{\tabcolsep}{4pt}
  \begin{tabular}{@{}lr cccc@{}}
    \toprule
    \textbf{Bug class} & \textbf{\#} & \textbf{Disc.} & \textbf{Triage} & \textbf{Gen.} & \textbf{Valid.} \\
    \midrule
    UAF            & 979 & \cmark & \cmark & \pmark & \pmark \\
    OOB            & 633 & \cmark & \cmark & \pmark & \pmark \\
    uninit (UBI)   & 497 & \cmark & \pmark & \pmark & \pmark \\
    null-deref     & 592 & \cmark & \pmark & \pmark & \pmark \\
    GPF            & 524 & \cmark & \pmark & \pmark & \pmark \\
    data-race      & 199 & \cmark & \pmark & \xmark & \xmark \\
    deadlock/lock  & 470 & \cmark & \pmark & \xmark & \xmark \\
    hang/stall     & 351 & \cmark & \pmark & \xmark & \xmark \\
    mem-leak       & 243 & \cmark & \cmark & \pmark & \pmark \\
    WARNING/other  & 2238 & \cmark & \pmark & \pmark & \xmark \\
    \bottomrule
  \end{tabular}
\end{table}

\edited{\Cref{tab:survey} lists representative systems for stages S2 to S5 with
their automation level, and \Cref{tab:discovery} gives the full discovery
classification.}

\begin{table}[!htb]
  \centering
  \caption{Representative techniques, stages S2--S5 (\Cref{tab:discovery}
  covers S1). \emph{Auto.}: A0 production-deployed, A1 offline/prototype, H
  human-in-loop, M manual. \ding{72}~LLM-based, (U)~user-space contrast.}
  \label{tab:survey}
  \scriptsize
  \setlength{\tabcolsep}{3.5pt}
  \renewcommand{\arraystretch}{0.90}
  \begin{tabular}{@{}lllc@{}}
      \toprule
      \textbf{System} & \textbf{Focus} & \textbf{Auto.} & \textbf{LLM} \\
      \midrule
            \multicolumn{4}{@{}l}{\textit{\stage{S2} Triage}} \\
      SyzScope~\cite{syzscope}     & impact re-rank  & A1 & \\
      LLM triage~\cite{llmtriage}  & severity        & A1 & \ding{72} \\
      GREBE~\cite{grebe}           & exploitability  & A1 & \\
      AURORA~\cite{aurora}         & root cause      & A1 & \\
      ARCUS~\cite{arcus}           & root cause      & A1 & \\
      \midrule
      \multicolumn{4}{@{}l}{\textit{\stage{S3} Generation}} \\
      CrashFixer~\cite{crashfixer}  & LLM agent      & A1 & \ding{72} \\
      PatchIsland~\cite{patchisland}& LLM agent      & A1 & \ding{72} \\
      kGym/kBench~\cite{kgym}        & platform       & A1 & \ding{72} \\
      Coccinelle~\cite{coccinelle}   & semantic       & A0 & \\
      FixMorph~\cite{fixmorph}       & backport       & A1 & \\
      PatchNet~\cite{patchnet}       & backport sel.  & A1 & \\
      \midrule
      \multicolumn{4}{@{}l}{\textit{\stage{S4} Validation}} \\
      KLAUS~\cite{klaus}             & correctness    & H  & \\
      LLM-judge~\cite{llmjudge}      & correctness (U)& H  & \ding{72} \\
      Incomplete-fix~\cite{incompletefix}& completeness& A1 & \\
      PDiff~\cite{pdiff}             & presence       & A1 & \\
      \midrule
      \multicolumn{4}{@{}l}{\textit{\stage{S5} Integration}} \\
      CodeReviewer~\cite{codereviewer}& review autom. & A1 & \ding{72} \\
      Zhou et al.~\cite{zhou2017}    & maintainer load& M  & \\
      Rust-for-Linux~\cite{rust4linux}& reviewer scarce& M & \\
      Patch-Me~\cite{patchmeifyoucan}& CVE flood      & M  & \\
      Jiang et al.~\cite{jiang2013}  & acceptance     & M  & \\
      Patchwork~\cite{patchwork}     & dataset        & M  & \\
      \bottomrule
  \end{tabular}
\end{table}

\edited{\myparagraph{\Cref{tab:discovery}} The Tech.\ column lists
each system's technique families, primary first. Dynamic families are
coverage or execution (Cov), input-structure or specification inference
(Inp), dependency or sequence inference (Dep), mutation or task scheduling
(Sch), state-aware fuzzing (Sta), directed fuzzing (Dir), concurrency
interleaving (Con), and device emulation (Emu). Static families are taint or
dataflow analysis (Tnt), typestate or lifecycle analysis (Typ), symbolic or
path-sensitive reasoning (Sym), specification or pattern mining (Min), and
learned models (ML).}

\begin{table}[!htb]
  \centering
  \caption{Full \stage{Discovery} (S1) classification. \emph{Type}: F=fuzzing,
  S=static, H=hybrid, E=empirical study. \emph{Bug class}: memory (mem),
  concurrency (conc), logic, semantic (sem), taint. \ding{72}~LLM-based.}
  \label{tab:discovery}
  \scriptsize
  \setlength{\tabcolsep}{1.3pt}
  \renewcommand{\arraystretch}{0.90}
  \begin{tabular}{@{}llll >{\raggedright\arraybackslash}p{0.78in} c@{}}
      \toprule
      \textbf{System} & \textbf{Type} & \textbf{Target} & \textbf{Bug class} & \edited{\textbf{Tech.}} & \textbf{LLM} \\
      \midrule
      \multicolumn{6}{@{}l}{\textit{Dynamic discovery (fuzzing)}} \\
    kAFL~\cite{kafl}            & F & generic     & mem & \edited{Cov} & \\
    Unicorefuzz~\cite{unicorefuzz} & F & generic   & mem & \edited{Cov} & \\
    Agamotto~\cite{agamotto}    & F & driver      & mem & \edited{Cov, Emu} & \\
    Horus~\cite{horus}          & F & generic     & mem & \edited{Cov} & \\
    BoKASAN~\cite{bokasan}      & F & generic     & mem & \edited{Cov} & \\
    MoonShine~\cite{moonshine}  & F & core        & mem & \edited{Dep, Cov, Tnt} & \\
    HEALER~\cite{healer}        & F & core        & mem & \edited{Dep, Cov} & \\
    ACTOR~\cite{actor}          & F & core        & mem & \edited{Sta, Dep, Tnt} & \\
    MOCK~\cite{mock}            & F & core        & mem & \edited{Dep, Sch, ML, Cov} & \\
    SyzVegas~\cite{syzvegas}    & F & core        & mem & \edited{Sch, Cov} & \\
    Snowplow~\cite{snowplow}    & F & core        & mem & \edited{Sch, ML, Dir, Cov} & \\
    StateFuzz~\cite{statefuzz}  & F & driver      & mem,logic & \edited{Sta, Sym, Sch} & \\
    HFL~\cite{hfl}              & H & core        & mem & \edited{Sym, Dep, Inp, Tnt, Sta} & \\
    CountDown~\cite{countdown}  & F & core        & mem (refcnt) & \edited{Sta, Dep, Sch} & \\
    Bin-Cov~\cite{liu2024leveraging} & F & generic & mem & \edited{Cov, Tnt} & \\
    DIFUZE~\cite{difuze}        & F & driver      & mem & \edited{Inp, Sym, Tnt} & \\
    IMF~\cite{imf}              & F & API (macOS) & mem & \edited{Dep, Inp, Min} & \\
    DR.~FUZZ~\cite{drfuzz}      & F & driver      & mem & \edited{Inp, Sta, Sch, Tnt, Cov} & \\
    PrIntFuzz~\cite{printfuzz}  & F & driver      & mem & \edited{Emu, Inp, Tnt, Sym, Cov} & \\
    KextFuzz~\cite{kextfuzz}    & F & drv (macOS) & mem & \edited{Cov, Inp, Tnt, Dep} & \\
    NTFUZZ~\cite{ntfuzz}        & F & drv (Win) & mem & \edited{Inp, Tnt, Sch} & \\
    JANUS~\cite{janus}          & F & FS          & mem,sem & \edited{Sta, Sch, Cov} & \\
    Hydra~\cite{hydra_sosp}       & F & FS     & sem & \edited{Sta, Sch, Cov} & \\
    Razzer~\cite{razzer}        & H & conc.       & conc & \edited{Con, Dir, Cov, Tnt} & \\
    SegFuzz~\cite{segfuzz}      & F & conc.       & conc & \edited{Con, Cov} & \\
    Snowboard~\cite{snowboard}  & F & conc.       & conc & \edited{Con, Sch, Dep} & \\
    SyzDirect~\cite{syzdirect}  & F & directed    & mem,logic & \edited{Dir, Dep, Inp, Sch, Tnt} & \\
    SyzRisk~\cite{syzrisk}      & F & regression  & mem,logic & \edited{Dir, Sch, Cov} & \\
    SyzGen~\cite{syzgen}        & F & spec-gen    & n/a & \edited{Inp, Dep, Sym, Cov, Min} & \\
    KSG~\cite{ksg}              & F & spec-gen    & n/a & \edited{Inp, Sym, Tnt} & \\
    SyzDescribe~\cite{syzdescribe} & S & spec-gen & n/a & \edited{Inp, Dep, Tnt} & \\
    SyzGen++~\cite{chen2024syzgen++} & F & spec-gen & n/a & \edited{Dep, Inp, Sym, Cov} & \\
    FuzzNG~\cite{bulekov2023no} & F & core        & mem & \edited{Inp, Cov} & \\
    KernelGPT~\cite{kernelgpt}  & F & spec-gen    & n/a & \edited{Inp, Dep, ML} & \ding{72} \\
    Enterprise~\cite{enterprise} & E & generic    & mem,conc,logic & \edited{Cov} & \\
    Directed-Ind.~\cite{directed_industry} & E & directed & mem,logic & \edited{Dir, Inp, Sch, Cov} & \\
      \midrule
      \multicolumn{6}{@{}l}{\textit{Static and hybrid discovery}} \\
    DR.~CHECKER~\cite{drchecker}& S & driver      & mem,logic & \edited{Tnt} & \\
    K-Miner~\cite{kminer}       & S & core        & mem & \edited{Tnt} & \\
    CRIX~\cite{crix}            & S & generic     & logic (chk) & \edited{Min, Tnt} & \\
    LRSan~\cite{lrsan}          & S & generic     & logic (chk) & \edited{Min, Tnt} & \\
    UBITect~\cite{ubitect}      & S & generic     & mem (UBI) & \edited{Sym, Tnt, Typ} & \\
    K-MELD~\cite{kmeld}         & S & modules     & mem (leak) & \edited{Min, Typ, Tnt, Sym} & \\
    Goshawk~\cite{goshawk}      & S & generic     & mem & \edited{Min, Typ, Tnt, Sym, ML} & \\
    DCUAF~\cite{dcuaf}          & S & driver      & conc (UAF) & \edited{Tnt, Min} & \\
    DEADLINE~\cite{xu2018precise} & S & generic  & conc & \edited{Sym, Tnt} & \\
    VulDeePecker~\cite{li2018vuldeepecker} & S & generic & mem & \edited{ML, Tnt} & \\
    CheQ~\cite{lu2019automatically} & S & generic & logic (chk) & \edited{Min, Tnt} & \\
    Uninit-Bin~\cite{garmany2019static} & S & binary & mem (uninit) & \edited{Sym, Tnt} & \\
    SUTURE~\cite{zhang2021statically} & S & generic & taint & \edited{Tnt, Sym} & \\
    Kconfig~\cite{oh2021finding} & S & config     & logic & \edited{Sym} & \\
    KUBO~\cite{liu2021kubo}     & S & generic     & mem (UB) & \edited{Sym, Tnt} & \\
    DEPA~\cite{zhong2020inferring} & S & generic  & logic & \edited{Min, Tnt} & \\
    MANTA~\cite{yang2022making} & S & container   & mem (acct) & \edited{Tnt} & \\
    IncreLux~\cite{zhai2022progressive} & S & generic & mem (UBI) & \edited{Sym, Tnt} & \\
    PATA~\cite{li2022path}      & S & generic     & mem,logic & \edited{Tnt, Typ, Sym} & \\
    UACatcher~\cite{ma2023top}  & S & driver      & conc (UAC) & \edited{Typ, Tnt, Sym} & \\
    Err-Spec~\cite{dossche2024inference} & S & generic & logic (err) & \edited{Min, Tnt, Sym} & \\
    LLift~\cite{li2024enhancing} & S & generic    & mem (UBI) & \edited{Sym, ML} & \ding{72} \\
    IMMI~\cite{liu2024detecting} & S & generic    & mem & \edited{Min, Typ, Tnt, ML} & \ding{72} \\
    BugLens~\cite{li2025towards} & S & generic    & taint & \edited{Tnt, Sym, ML} & \ding{72} \\
    UAFX~\cite{zhang2025statically} & S & generic & mem (UAF) & \edited{Typ, Tnt, Sym} & \\
    SCAD~\cite{man2025scad}     & H & network     & sem & \edited{Sym, Tnt} & \\
    CPG~\cite{yamaguchi2014modeling} & S & generic & mem,logic & \edited{Min, Tnt} & \\
    Coccinelle~\cite{coccinelle}& S & generic     & logic & \edited{Min} & \\
    KNighter~\cite{knighter}    & S & generic     & logic & \edited{Min, ML, Sym} & \ding{72} \\
      \bottomrule
  \end{tabular}
\end{table}

\myparagraph{Threats to validity.} Our measurement studies the fixed
population from one (dominant) source, syzbot. \edited{Bugs reported elsewhere
often arrive with a patch attached, skipping the interval we measure, so our
figures characterize the fuzzer-found population,} and right-censoring
understates latency, both conservative with respect to our thesis.
\edited{Two authors independently labeled all 110 systems. Agreement was
88.2\% ($\kappa = 0.81$) for stage, 81.8\% ($\kappa = 0.72$) for challenge,
and 83.6\% ($\kappa = 0.75$) for automation level. Because a system can
carry several challenge labels, a system counts as agreed on challenge only
when both authors assign the same set of labels, and $\kappa$ is computed on
that set-level judgment. Disagreements concentrated on the A1 versus H
boundary and were resolved by discussion, with the stricter label chosen
when a tool proposes but a human decides. Reassigning every disputed label,
whether stage, challenge, or the A1 versus H boundary, moves no technique
across the front-end and back-end divide, so neither the automation
gradient nor the coverage-gap matrix changes.} Patch-series reconstruction from
mailing archives is incomplete, so revision and review-effort figures are lower
bounds, and $\sim$35\% of our headline latency is post-merge confirmation lag
rather than engineering time (\Cref{sec:meas-decomp}). \edited{Our benchmark and LLM-assisted labeling face the pitfalls of
LLM-based security evaluation catalogued by Evertz et al.~\cite{llmpitfalls},
including training-data contamination, prompt sensitivity, reliance on an LLM
judge, and absent execution feedback. We mitigate these with recorded model versions, two judge models with author
re-judging of disagreements and an independent author check on 100 sampled
candidates, and by releasing every verdict and
rationale for audit (\Cref{sec:gen-bench}). On the buildable subset, the
judge's verdicts matched real rebuild-and-reproduce outcomes in 32 of 36
cases.} A different
lens on the
same papers is possible.

\myparagraph{Open science and ethics.}
We release the full dataset of 6{,}946 syzbot-fixed Linux kernel bugs, the
analysis scripts behind every figure and statistic, and the classification of
the 140 surveyed papers. \edited{We will also release the judge rubric,
verdicts, and rationales from \Cref{sec:gen-bench}.} This work studies only public data on already-patched
defects and reports only aggregate statistics.


\end{document}